\documentclass[pra,showpacs,superscriptaddress,nofootinbib,notitlepage,amssymb]{revtex4-1}
\usepackage{graphicx}
\usepackage{amsmath}
\usepackage{bbold}
\usepackage{amssymb}
\usepackage[usenames]{color}
\usepackage{hyperref}
\usepackage{natbib}
\usepackage{bm}
\usepackage{threeparttable}
\usepackage{subfigure}
\usepackage{upgreek }

\newcommand{\beq}{\begin{equation}}
\newcommand{\eeq}{\end{equation}}
\newcommand{\beqnn}{\begin{equation*}}
\newcommand{\eeqnn}{\end{equation*}}
\newcommand{\bea}{\begin{eqnarray}}
\newcommand{\eea}{\end{eqnarray}}
\newcommand{\beann}{\begin{eqnarray*}}
\newcommand{\eeann}{\end{eqnarray*}}
\newcommand{\bes} {\begin{subequations}}
\newcommand{\ees} {\end{subequations}}

\newcommand{\av}[1]{\langle #1\rangle}

\newcommand{\ket}[1]{ | #1\rangle}
\newcommand{\bra}[1]{\langle #1 | }

\newcommand{\ketbra}[2]{|#1\rangle\langle #2|}

\newcommand{\mE}{\mathcal{E}}
\newcommand{\mP}{\mathcal{P}}
\newcommand{\mQ}{\mathcal{Q}}

\newcommand{\mG}{\mathrm{G}}
\newcommand{\ident}{\mathbf{1}}

\newcommand{\Tr}{\mathrm{Tr}}

\newcommand{\ignore}[1]{}

\begin{document}
\title{Fluctuation theorems for quantum processes}
%\author{Tameem Albash,$^{(1,2)}$, Daniel A. Lidar,$^{(1,2,3,4)}$, Milad Marvian$^{(2,3)}$, Paolo Zanardi$^{(1,2)}$}
%%\affiliation{$^{(1)}$Department of Physics and Astronomy, $^{(2)}$Center for Quantum Information Science \&
%%Technology, $^{(3)}$Information Sciences Institute, $^{(4)}$Department of Electrical Engineering, $^{(5)}$Department of Chemistry\\University of Southern California, Los Angeles, California
%%90089, USA}
%\address{$^1$ Department of Physics and Astronomy}
%\address{$^2$ Center for Quantum Information Science \&
%Technology}
%\address{$^3$ Department of Electrical Engineering}
%\address{$^4$ Department of Chemistry\\University of Southern California, Los Angeles, California
%90089, USA}
%\eads{\mailto{albash@usc.edu}, \mailto{lidar@usc.edu}}
\author{Tameem Albash}
\affiliation{Department of Physics and Astronomy}
\affiliation{Center for Quantum Information Science \&
Technology}
\author{Daniel A. Lidar}
\affiliation{Department of Physics and Astronomy}
\affiliation{Center for Quantum Information Science \&
Technology}
\affiliation{Department of Electrical Engineering}
\affiliation{Department of Chemistry\\University of Southern California, Los Angeles, California
90089, USA}
\author{Milad Marvian}
\affiliation{Center for Quantum Information Science \&
Technology}
\affiliation{Department of Electrical Engineering}
\author{Paolo Zanardi}
\affiliation{Department of Physics and Astronomy}
\affiliation{Center for Quantum Information Science \&
Technology}

\begin{abstract}
We present fluctuation theorems and moment generating function equalities for generalized thermodynamic observables and quantum dynamics described by completely positive trace preserving (CPTP) maps, with and without feedback control. Our results include the quantum Jarzynski equality and Crooks fluctuation theorem, and clarify the special role played by the thermodynamic work and thermal equilibrium states in previous studies. We show that for a specific class of generalized measurements, which include projective measurements, unitality replaces microreversibility as the condition for the physicality of the reverse process in our fluctuation theorems. We present an experimental application of our theory to the problem of extracting the system-bath coupling magnitude, which we do for a system of pairs of coupled superconducting flux qubits undergoing quantum annealing.
\end{abstract}

%\submitto{\NJP}
\maketitle

%
%%%%%%%%%%%%%%%%%%%%%%
\section{Introduction}
%%%%%%%%%%%%%%%%%%%%%%%
Fluctuation theorems provide powerful analytical tools for nonequilibrium physics.  Some $15$ years ago Jarzynski discovered an equality for classical processes that shows how to determine free energy changes by measuring only the work performed on the system, without the need to determine the accompanying entropy changes, and in particular without the requirement that the processes be quasistatic. 
Consider for example two thermal equilibrium states $A$ and $B$ of a system, each state with different macroscopic thermodynamic observables such as pressure and volume, but both at a fixed inverse temperature $\upbeta$.  The system is initially in state $A$, and work is performed on the system according to some protocol to drive it to the macroscopic conditions of state $B$.  If the system is not allowed to equilibrate, the system may not reach the thermal equilibrium state $B$.  Nevertheless, for this forward process, the classical Jarzynski equality (CJE)
\cite{1997PhRvL..78.2690J}
\begin{equation} 
\label{eqt:Jarzynski}
\langle e^{-\upbeta (w-\Delta F)} \rangle = \gamma \ , 
\end{equation}
where $\gamma =1$, relates the statistical average $\av{\,}$ of the work $w$ done on the driven system to the free energy difference $\Delta F$ of the final equilibrium state $B$ (whether this state is reached or not) and the initial thermal equilibrium (Gibbs) state $A$.  In particular, this result is independent of what protocol is used, which is one of its remarkable features.  In the presence of feedback control (``Maxwell's demon") the efficacy parameter $\gamma$ differs from unity, and characterizes the efficacy of feedback and the amount of information extracted \cite{2010PhRvL.104i0602S}.

The CJE follows directly from the Tasaki-Crooks fluctuation theorem \cite{PhysRevE.60.2721,2000cond.mat..9244T}, which relates the probability density function (PDF) of work done in the forward process $P_F(w)$ to the PDF of a reverse process $P_R(w)$:
\begin{equation} 
\label{eqt:Crooks}
{P_F(w)} e^{-\upbeta( w-\Delta F)} = {P_R(-w)} \ .
\end{equation}
The reverse process describes the evolution of the system starting from the thermal equilibrium state $B$ and applying an appropriately time-reversed work protocol on the system, although this may not correspond to the time-reversed evolution of the forward process.  A key element of the fluctuation theorem is the requirement of a microreversibility condition, which relates the forward and reverse dynamics at any given instant in time.  For example, for driven classical systems, microreversiblity relates the flow of phase space points under the forward driving protocol to the flow under the reversed driving protocol, via the heat absorbed by the system \cite{stratonovich1994nonlinear,RevModPhys.83.771}. For a specific pertinent statement of microreversiblity see, e.g., Eq.~(5) in Ref.~\cite{PhysRevE.60.2721}.  Many generalizations have been developed (for reviews see, e.g., \cite{PT:05,doi:10.1146/annurev-conmatphys-062910-140506,RevModPhys.83.771}), with appropriate generalizations of the imposed microreversiblity condition. In particular, the classical results (see also \cite{Seifert:05}) have been quantized, first for thermal states undergoing unitary evolution \cite{2000cond.mat..7360K,2000cond.mat..9244T}, and subsequently for thermal states undergoing non-unitary, open system dynamics \cite{JPSJ.69.2367,PhysRevLett.90.170604,2005PhRvE..72b7102M,PhysRevE.75.050102,1742-5468-2008-10-P10023,2008arXiv0812.4955Q,2009PhRvL.102u0401C,2009JSMTE..02..025T,2011PhRvL.107n0404D},
including continuous monitoring \cite{PhysRevLett.105.140601} and quantum feedback \cite{Sagawa:08,springerlink:10.1007/s10955-011-0153-7,2010PhRvL.104i0602S}.  
 
Here we aim to show that there exists a single unified framework from which all the quantum results can be derived as well as generalized, using only basic tools of the theory of open quantum systems \cite{Breuer:book} and quantum information theory \cite{Nielsen:book}.
To this end we derive a general fluctuation theorem for quantum processes described by completely positive trace-preserving (CPTP) maps $\mE$, with or without feedback. CPTP maps arguably represent the most general form of open quantum system dynamics, under the assumption of an initially uncorrelated system-bath state \cite{Breuer:book}.
% since any CPTP map can be shown to arise from a larger Hilbert space undergoing unitary dynamics \cite{Stinespring}.  
Our strategy leads to a general and simple recipe for writing down fluctuation theorems, not all of which must correspond to a measurable thermodynamic observable (note that work is not a quantum observable \protect\cite{PhysRevE.75.050102}), or involve a physical reverse process. 
We show that in order for a PDF for the reverse process to exist, the map $\mE$ must be \emph{unital}, and we show that the map describing the reverse process is simply the dual map $\mE^{\ast}$ of the forward process, which leaves no ambiguity in defining the reverse process.  In this sense unital channels emerge as playing a crucial role in fluctuation theory for any CPTP map, replacing the role typically played by the standard thermodynamic notion of microreversibility.  Our work illuminates the special role played by the Gibbs state and work measurements, both of which feature prominently in the literature on fluctuation theorems.

We empirically determine the first moment of our integral fluctuation theorem in our theory via an experiment involving pairs of superconducting flux qubits on a programmable chip.  The first moment turns out to be a measure of the information-geometric distance of the evolved state and the virtual final equilibrium state, where virtual here signifies that the equilibrium state is never actually reached by our evolution.  We show that these experimental results can be well explained using a time-dependent Markovian master equation with a free adjustable parameter determining the system-bath coupling strength. As a novel application, our theory provides a meaningful optimization target that allows us to determine this parameter by fitting to the experimental data. We thus establish quantum fluctuation theorems as important tools for studying open quantum systems.  

%%%%%%%%%%%%%%%%%%%%%%
\section{General Fluctuation Theorem}
\subsection{Review of Quantum Jarzynski Equality for Closed Systems with Projective Measurements}
\label{sec:closed-QJE}
We briefly review the generalization of the CJE to the quantum case of thermal states undergoing unitary evolution \cite{2000cond.mat..9244T}, as it will help set the stage for our work (see also \ref{app:D}).  We consider a Hamiltonian $H(t)$ that interpolates between two system states described by $H(0)$ and $H(t_f)$, with an associated unitary time evolution operator $U(t_f,0) = T_+ \exp \left(-i \int_0^{t_f} H(t) dt \right)$.  The system, described by a density matrix $\rho(t)$, is initially in the Gibbs state associated with $H(0)$:
\beq
\rho(0) = \frac{1}{Z(0)} e^{- \upbeta H(0)} \ ,
\eeq
where $Z(t)$ is the partition function associated with $H(t)$.  A (projective) measurement $\mathcal{P} := \left\{ P_{\alpha} =  \ketbra{\varepsilon_\alpha(0)}{\varepsilon_\alpha(0)}  \right\}$ of the eigenenergy (associated with $H(0)$) is performed, which selects the energy state $| \varepsilon_\alpha(0) \rangle$ with probability $p_\alpha = e^{- \upbeta \varepsilon_\alpha(0)}/Z(0)$.  The system is then evolved according to $U(t_f,0)$, and another projective measurement $\mathcal{Q} := \left \{Q_\beta =  \ketbra{\varepsilon_\beta(t_f)}{\varepsilon_\beta(t_f)} \right\}$ of the eigenenergy (associated with $H(t_f)$) is performed.  The conditional probability of measuring the energy $\varepsilon_\beta(t_f)$ is given by:
 \beq
p_{\beta|\alpha} = \Tr \left[ Q_\beta U(t_f,0) P_{\alpha} U^\dagger(t_f,0) \right] \ .
 \eeq
Let us define the work performed during this evolution by $W_{\alpha \beta} = \varepsilon_\beta(t_f) - \varepsilon_\alpha(0)$ (this definition applies since the system is closed).  The PDF associated with $W$ is:
 \beq
 P_F(w) = \sum_{\alpha, \beta} \delta (w - W_{\alpha \beta}) p_{\beta | \alpha}  p_{\alpha} \ .
 \eeq
Using that $\upbeta \Delta F =  \ln(Z(0)/Z(t_f))$ and the Dirac delta properties $\delta(x-x_0)f(x) = \delta(x-x_0)f(x_0)$, and $\delta(x) = \delta(-x)$, we note that upon multiplying both sides of the equation by $e^{-\upbeta \left( w - \Delta F \right)}$, we have
 \bes 
 \begin{align}
\hspace{-1.5cm} P_F(w) e^{-\upbeta \left( w - \Delta F \right)} &=  \sum_{\alpha \beta} \delta(-w - \tilde{W}_{\beta \alpha }) \Tr \left[P_{\alpha} U^\dagger (t_f,0) Q_\beta U(t_f,0) \right] \frac{e^{- \upbeta \varepsilon_\beta (t_f)}}{Z(t_f)} \ , \\
 &= P_R(-w) \label{eqt:closedFluctuation}\ ,
 \end{align}
\ees
where we have identified the right hand side with the PDF of the reverse process with work $\tilde{W}_{\beta \alpha} = - W_{\alpha \beta}$, for which the following temporally ordered sequence applies: (i) the system is initially in the Gibbs state associated with $H(t_f)$, (ii) a projective measurement $\mathcal{Q}$ is performed with outcome probability $q_\beta = e^{- \upbeta \varepsilon_\beta(t_f)}/Z(t_f)$, (iii) the system evolves unitarily via $U^{\dagger}(t_f,0)$, and (iv) a projective measurement $\mathcal{P}$ is performed.  Eq.~\eqref{eqt:closedFluctuation} is the closed system fluctuation theorem, the integration of which gives a closed system quantum Jarzynski equality exactly of the form of Eq.~\eqref{eqt:Jarzynski}. Note that in this case the reverse process is simply the time-reversed process, a situation that will change in our more general analysis below.
%
%%%%%%%%%%%%%%%%%%%%%%
\subsection{Quantum Jarzynski Equality for Generalized Measurements}
%%%%%%%%%%%%%%%%%%%%%%
%
We now generalize the previous section result beyond thermal states, unitary evolutions, and projective measurements.  Consider a fiducial initial state $\rho$,
two sets of generalized measurements $\mP:=\{P_{\alpha}\}$ and $\mQ:=\{Q_{\beta}\}$ (see also \cite{springerlink:10.1007/s10955-011-0153-7}), a CPTP map $\mE$ (see also \cite{2012PhRvA..86d4302K, Vedral:2012}), and a fixed, yet arbitrary distribution $q:=\{q_\beta\}$, whose role we clarify later.
%.Although arbitrary at this point, the distribution $q$ will be shown to determine the observable (later called $V$) in our fluctuation theorem and the initial state of the appropriately defined reverse process.  
The quintuple $(\rho,\mP,\mE,\mQ,q)$ is the basic input data describing the problem. The measurement operators $P_{\alpha}$ satisfy $\sum_\alpha P_{\alpha}^\dag P_{\alpha} = \ident$ (the identity operator), and similarly for $\mQ$. Generalized measurements are the most general kind of measurements in quantum theory, and they include projective measurements, positive operator valued measures (POVM), and weak measurements as special cases \protect{\cite{Nielsen:book,PhysRevLett.95.110409}}.  The CPTP map $\mE$ has Kraus operators $\{A_i\}$, i.e., 
\beq
\mE(X) = \sum_i A_i X A_i^{\dagger} \ ,
\eeq
where $\sum_i A_i^{\dagger}A_i =\ident$.  

\begin{figure}[t] %
   \centering
   \includegraphics[width=2.8in]{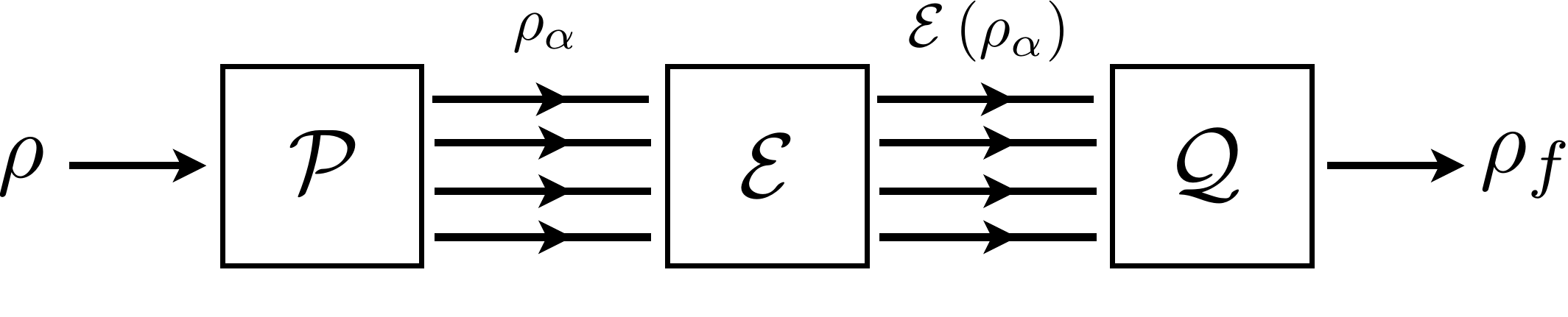} 
   \caption{The forward process protocol. A quantum state $\rho$ is prepared, measured ($\mP$), evolved via a CPTP map $\mE$, and measured again ($\mQ$).}
   \label{fig:Protocol1}
\end{figure}

We first consider the \emph{forward} process depicted in Fig.~\ref{fig:Protocol1}.
A mixed state ensemble $\rho_p = \sum_\alpha p_\alpha \rho_\alpha$ is prepared by measuring $\mP$, so that the normalized state 
\beq
\rho_{\alpha} = P_{\alpha}\rho P_{\alpha}^\dag/p_\alpha
\eeq 
has probability $p_\alpha = \Tr[P_{\alpha}^\dag P_{\alpha}\rho]$\footnote{For a given generalized measurement $\mathcal{P}$, the set of possible values of the probabilities $p_{\alpha}$ is in general constrained, even if the initial state $\rho$ is arbitrary (see \ref{app:A1} for an example of this).}.  Next $\rho_\alpha$ evolves under $\mE$, and finally the measurement $\mQ$ is performed.  The conditional probability of observing outcome $\beta$ given outcome $\alpha$ is then
\beq
p_{\beta|\alpha} = \Tr[Q_\beta^\dag Q_\beta {\mE}(\rho_{\alpha})] \ .
\eeq
The marginal probability distribution of outcomes is $f:=\{f_\beta\}$, where
\beq
f_\beta = \sum_{\alpha} p_{\beta|\alpha}p_\alpha = \sum_{\alpha} p_{(\alpha, \beta)} \ ,
\eeq
and where $p_{(\alpha, \beta)}$ is the joint probability distribution.  In the last equality we used Bayes' rule for the joint probability $p_{(\alpha, \beta)}$.  We therefore have:
\beq
f_\beta  = \Tr\left[Q_\beta^\dag Q_\beta \mE(\rho_p)\right] \ .
\eeq
%
%where $\tilde{\rho}_{\beta}:=Q_\beta^\dag Q_\beta$.
%We next define the (virtual) final state ${\rho}_q := \sum_{\beta=1}^d q_\beta \tilde{\rho}_{\beta}$, where $\tilde{\rho}_{\beta}:=Q_\beta^\dag Q_\beta >0$ is a normalized state having probability $q_\beta>0$ $\forall \beta$. Note that $d=\Tr[\ident]$ is the system Hilbert space dimension and that $f_\beta  = \Tr\left[\tilde{\rho}_{\beta} \mE(\rho_p)\right]$.
Note that the transition matrix $M:=\{p_{\beta|\alpha}\}_{\alpha,\beta}$ of the forward process is column stochastic:
\beq
\sum_\beta p_{\beta|\alpha} =  \Tr\left[\sum_\beta Q_\beta^\dag Q_\beta {\mE}(\rho_{\alpha})\right] = \Tr[ {\mE}(\rho_{\alpha})] = \Tr[ \rho_{\alpha}] = 1\ ,
\label{eq:colsto}
\eeq
where we used the normalization condition of the generalized measurement $\mathcal{Q}$ and the fact that $\mE$ is a trace preserving map.

 Let the random variable $V$ be a real-valued function parametrized by the measurement outcomes $\{\alpha, \beta\}$. $V$ will play the role of a generalized thermodynamic observable, where we use the term `observable' in a loose sense since it is typically an abstract quantity and only is a thermodynamic observable in special cases.  The PDF associated with $V$ is
\bes
\label{eq:P}
\begin{align}
\label{eq:3a}
P_{\mE} (v)&:=\sum_{\alpha, \beta} \delta(v -V_{\alpha \beta}) p_{(\alpha, \beta)}  \\
\label{eq:3b}
&= \sum_{\alpha, \beta} \delta(v -V_{\alpha \beta}) \Tr[ Q_\beta^\dag Q_\beta
 {\mE}(\rho_{\alpha})] p_\alpha \ .
\end{align}
\ees
Let us now choose 
\beq
V_{\alpha \beta}=\ln( p_\alpha/q_\beta)\, ,
\label{eq:V_choice}
\eeq 
(note that such a form will give us the expression found in Eq.~\eqref{eqt:closedFluctuation}) and also define the generalized reverse thermodynamic observable $\tilde{V}_{\beta \alpha} := - V_{\alpha \beta} = \ln(q_\beta/ p_\alpha )$.  This choice of $V_{\alpha \beta}$ requires that $p_{\alpha} \neq 0 \  \forall \alpha , \ q_{\beta} \neq 0 \ \forall  \beta$.  Then, using the Dirac delta properties $\delta(x-x_0)f(x) = \delta(x-x_0)f(x_0)$, and $\delta(x) = \delta(-x)$, we find
\bes \label{eqt:6}
\begin{align}
P_{\mE}(v) e^{-v} & = \sum_{\alpha, \beta} \delta(v -V_{\alpha \beta}) \Tr[\tilde{\rho}_{\beta} {\mE}(\rho_{\alpha})]  q_\beta \ , \label{eq:4a}
\\
\label{eq:4b}
&= \sum_{\alpha, \beta} \delta (-v - \tilde{V}_{\beta \alpha}) \Tr[\rho_{\alpha}{\mE}^{\ast}(\tilde{\rho}_{\beta}) ]   q_\beta =: \mathcal{F}_{\mathcal{E}^\ast}(-v) \ ,
\end{align}
\ees
where $\tilde{\rho}_{\beta} : = Q_\beta^\dag Q_\beta$ and
where we used the dual map $\mE^{\ast}$ with Kraus operators $\{A_i^\dag\}$, i.e., $\mE^{\ast}(X) = \sum_i A_i ^\dag X A_i$,  for which $ \Tr[A{\mE}(B)] = \Tr[B{\mE}^{\ast}(A)]$ for any pair of operators $A$ and $B$.  

Comparing Eqs.~\eqref{eq:3b} and \eqref{eq:4b}, it is tempting to identify the latter with a PDF $\tilde{P}_{\mathcal{E}^{\ast}}(-v)$ associated with the dual map ${\mathcal{E}^{\ast}}$, however there are important differences.  
First, while the map $\mathcal{E}$ acts on a normalized state $\rho_{\alpha}$, the ``state" $\tilde{\rho}_{\beta} $ acted upon by dual map is not necessarily normalized.  Second, while $\sum_{\beta} \tilde{\rho}_{\beta} = \sum_{\beta} Q_\beta^{\dagger} Q_{\beta} = \ident$, so that the set $\{\tilde{\rho}_{\beta}\}$ forms a POVM, $\sum_{\alpha} \rho_{\alpha}$ is not necessarily equal to the identity operator, so the set $\{\rho_{\alpha}\}$ cannot always be identified with a POVM.  For this reason we have for the time being used the notation $\mathcal{F}_{\mathcal{E}^\ast}(-v)$ in Eq.~\eqref{eq:4b}. We revisit this issue in subsection~\ref{sec:fluctuation}, where we show under which conditions $\mathcal{F}_{\mathcal{E}^\ast}(-v)$ can be interpreted as the PDF of a reverse process.

We define the \emph{efficacy}\footnote{We use the term efficacy loosely here.  In the case of the classical Jarzynski equality with feedback \cite{2010PhRvL.104i0602S}, the right hand side of the equality is indeed a measure of the efficacy of the feedback protocol.  Here we make no such claims.} $\gamma$ \cite{2010PhRvL.104i0602S,2012PhRvA..86d4302K} as 
\beq
\gamma : = \int_{-\infty}^{\infty} \hspace{-0.25cm} dv\, \mathcal{F}_{\mathcal{E}^\ast}(v)= \sum_{\alpha,\beta} q_\beta  \Tr[ \tilde{\rho}_{\beta} {\mE}(  {\rho_{\alpha}} )] = \sum_{\alpha,\beta} \Tr [ \rho_{\alpha} {\mE}^{\ast} ( q_\beta \tilde{\rho}_{\beta} )] \ .
\label{eq:gamma}
\eeq
%Note that $\tilde{P}_{{\mE}^{\ast}}$ need not be a probability distribution (it is not necessarily normalized). 
Upon integration of Eq.~\eqref{eq:4a}, we arrive at what we call the quantum Jarzynski equality (QJE), as it generalizes the CJE, Eq.~\eqref{eqt:Jarzynski}:
\begin{equation}
\av{e^{-v}} = \int_{-\infty}^{\infty} dv \, P_{\mE}(v) e^{-v} =   \gamma \ .
\label{eqt:albash}
\end{equation}
If instead of the choice made in Eq.~\eqref{eq:V_choice} we choose $V_{\alpha\beta} = \ln \left( p_{\beta | \alpha} / q_\beta \right)$ \cite{Vedral:2012},  we find
\beq
P_\mathcal{E}(v)e^{-v} = \sum_{\alpha, \beta} \delta(v - V_{\alpha \beta}) p_{\alpha} q_\beta \ , 
\eeq
which upon integration gives the QJE with $\gamma=1$.
%, thus the choice \eqref{eq:V_choice} is more general

Using Jensen's inequality we have $\av{e^f} \geq e^{\av{f}}$ and thus find a generalized 2nd law of thermodynamics (we clarify this claim below):
\begin{equation}
\langle v\rangle \ge -\ln\gamma \ .
\label{II}
\end{equation}

We can substantially generalize the QJE Eq.~\eqref{eqt:albash} in terms of the moment generating functions for the map ${\mE}$ and its dual,
\beq
\chi_{\mE}(\lambda) := \int_{-\infty}^{\infty} d v P_{\mE}(v) e^{\lambda v}\, ,\quad \tilde{\chi}_{\mE^\ast}(\lambda) := \int_{-\infty}^{\infty} d v \mathcal{F}_{\mE^{\ast}}(v) e^{\lambda v} \ .
\label{eq:chidef}
\eeq
Multiplying Eq.~\eqref{eq:4a} by $e^{\lambda v}$ and integrating, we find:
\begin{eqnarray}
\chi_{\mE}(\lambda - 1) =
\tilde{\chi}_{\mE^{\ast}}(-\lambda) \ .
\label{eq:chi}
\end{eqnarray}
This extends the integral fluctuation relation Eq.~\eqref{eqt:albash} to all moments of the PDF $P_{\mE} (v)$. For example, setting $\lambda =0$, we recover the QJE Eq.~\eqref{eqt:albash}:
\beq
\chi_{\mE}(-1) = \langle e^{-v} \rangle = \tilde{\chi}_{\mE^{\ast}}(0)  = \gamma \ .
\eeq
Moreover, using $\langle v \rangle = \left. \frac{d}{d \lambda} \chi_{\mE}(\lambda) \right|_{\lambda = 0}$ we find (details can be found in \ref{app:A})
\beq
\langle v \rangle  = H(f\|q) + H(f) - H(p) \ ,
\label{eq:ent-gen-meas}
\eeq
where $H(f\| q) = -\sum_\beta f_\beta \ln(q_\beta)-H(f)$ is the relative entropy (Kullback-Leibler divergence), and $H(f) = -\sum_\beta f_\beta \ln(f_\beta)$ is the Shannon entropy. 

%
%%%%%%%%%%%%%%%%%%%%%%%%%%%%%%%%%%%%
\subsection{Fluctuation Theorem for ``Microreversible" Generalized Measurements and Unital Maps} \label{sec:fluctuation}
%%%%%%%%%%%%%%%%%%%%%%%%%%%%%%%%%%%%
%
Recall that in the discussion immediately following Eq.~\eqref{eqt:6} we stressed that $\mathcal{F}_{\mathcal{E}^\ast}(-v)$ cannot always be interpreted as the PDF associated with the dual map ${\mathcal{E}^\ast}$. Let us now restrict ourselves to a class of generalized measurements $\mathcal{P}$ and $\mathcal{Q}$  that satisfy additional constraints so that such an interpretation becomes possible:
\beq \label{eqt:constraint}
 \sum_{\alpha} \rho_{\alpha} = \ident\ \forall \rho \ , \quad \mathrm{Tr} \left[Q_{\beta}^\dag Q_{\beta} \right] = 1 \ .
 \eeq
 We call generalized measurements $\mathcal{P}$ and $\mathcal{Q}$ that satisfy Eq.~\eqref{eqt:constraint} ``microreversible" for reasons that will shortly become apparent.  Rank-$1$ projective measurements trivially satisfy these constraints, but are not necessary\footnote{An example of generalized measurements, which are not rank-1 projective measurements, satisfying Eq.~\eqref{eqt:constraint} are $P_1 = \sigma_+ = \frac{1}{2} \left(\sigma_x + i\sigma_y \right)$ and $P_2 = \sigma_-= \frac{1}{2} \left(\sigma_x - i\sigma_y \right)$ and $Q_1 = \sigma_x / \sqrt{2} , Q_2 = \sigma_y / \sqrt{2}$.}.  
 
Let $d = \Tr\left[\ident \right]$ denote the Hilbert space dimension; we prove in \ref{app:A2} that if the constraints \eqref{eqt:constraint} are satisfied then $\alpha, \beta \in \{1, \dots, d\}$, i.e., that none of the probabilities $p_\alpha$ or $q_\beta$ can vanish.  With these additional constraints, $\tilde{\rho}_{\beta}=Q_{\beta}^\dag Q_{\beta}$ can be identified as a normalized state, and we can define a new measurement $\tilde{\mathcal{Q}} := \left\{\tilde{Q}_\beta \right\}$ such that:
\beq \label{eqt:tildeQ}
\tilde{\rho}_{\beta} = \frac{ \tilde{Q}_\beta \tilde{\rho} \tilde{Q}_\beta^{\dag}}{q_\beta} \ , \quad Q_{\beta} = \tilde{U}_{\beta} \frac{ \sqrt{\tilde{\rho}} \tilde{Q}_{\beta}^{\dagger} }{\sqrt{q_\beta}} \ ,
\eeq
where $\tilde{U}_{\beta}$ is an arbitrary unitary operator, $\tilde{\rho}$ is a virtual final state, and the probability $q_\beta$ now takes the value $q_\beta = \Tr \left[ \tilde{Q}_\beta \tilde{\rho} \tilde{Q}_\beta^{\dag} \right]$ such that the mixed state ensemble $\rho_q = \sum_{\beta} q_\beta \tilde{\rho}_{\beta}$ is generated by measuring $\tilde{\mathcal{Q}}$ \footnote{Since the input data specifies ${\mathcal{Q}}$ rather than $\tilde{\mathcal{Q}}$, it is more natural to think of each measurement operator $Q_\beta$ as specifying the corresponding $\tilde{Q}_\beta$, i.e., $\tilde{Q}_\beta = \sqrt{q_\beta} Q_\beta^\dagger \tilde{U}_\beta (\tilde{\rho})^{-1/2}$.  The virtual final state $\tilde{\rho}$ should then be full rank in order for its inverse to exist.}.  Similarly, we can write the state $\rho_{\alpha}$ in terms of a new generalized measurement $\tilde{\mathcal{P}} := \left\{\tilde{P}_\alpha \right\}$:
\beq \label{eqt:tildeP}
\rho_{\alpha} = \tilde{P}_{\alpha}^{\dag} \tilde{P}_{\alpha} \ , \quad \tilde{P}_{\alpha} = U_{\alpha} \frac{\sqrt{\rho} P_{\alpha}^{\dagger}}{\sqrt{p_{\alpha}}}  \ .
\eeq
where ${U}_{\alpha}$ is an arbitrary unitary operator.  Thus 
\beq
\Tr[\tilde{\rho}_{\beta}]  =1 \quad \textrm{and} \quad 
\sum_\alpha \rho_\alpha = \ident\ .
\eeq 
Comparing with Eq.~\eqref{eq:3b}, we see that now $\mathcal{F}_{\mathcal{E}^\ast}(-v)$ [Eq.~\eqref{eq:4b}] can be identified with $\tilde{P}_{\mE^\ast}(-v)$, 
\beq \label{eqt:reverse}
\tilde{P}_{\mE^\ast}(v) = \sum_{\alpha \beta} \delta \left( v- \tilde{V}_{\beta \alpha} \right) \Tr[\tilde{P}_{\alpha}^\dagger \tilde{P}_{\alpha} {\mE}^{\ast}(\tilde{\rho}_{\beta}) ]   q_\beta \ ,
\eeq
associated with the dual map acting on the state $\tilde{\rho}_{\beta}$, followed by the generalized measurement $\tilde{\mathcal{P}}$. We have therefore arrived at our \emph{fluctuation theorem for CPTP maps}:
\beq
\label{eqt:CP-Crooks}
{P_{\mE}(v)} e^{-v} = {\tilde{P}_{\mE^\ast}(-v)}\ ,
\eeq
now bearing an obvious similarity to the Tasaki-Crooks fluctuation theorem, Eq.~\eqref{eqt:Crooks}.  Integrating this expression, we obtain (see also \cite{2012PhRvA..86d4302K})
\beq
\langle e^{-v} \rangle = \gamma = \Tr \left[\rho_q \mE(\ident) \right]= \Tr \left[\mE^{\ast}(\rho_q)\right] \ .
\eeq
A bound on the value of $\gamma$ is presented in \ref{app:L}.

\begin{figure}[t] %
   \centering
   \includegraphics[width=2.8in]{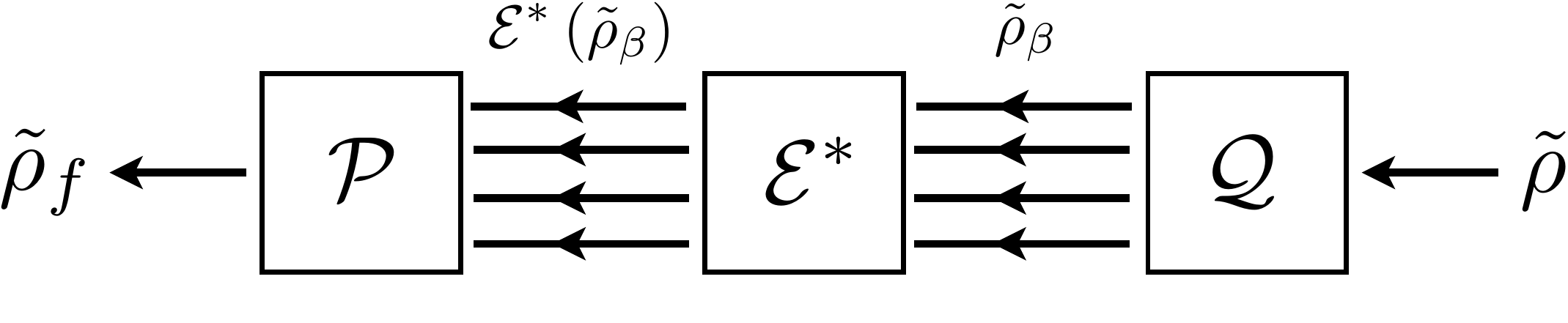} 
   \caption{The reverse process protocol. A quantum state $\tilde{\rho}$ is prepared, measured ($\tilde{\mQ}$), evolved via the dual CPTP map $\mE^\ast$, and measured again ($\tilde{\mP}$). The final state $\tilde{\rho}_f$ thus obtained is in general different from the state $\rho_p$.}
   \label{fig:ReverseProtocol}
\end{figure}

One more condition must be imposed in order for $\tilde{P}_{{\mE}^{\ast}}$ to become a PDF, i.e., for $\gamma$ [Eq.~\eqref{eq:gamma}] to equal $1$, namely, ${\mE}^{\ast}$ should itself be a CP map. This is the case if $\mE$ is unital [$\mE(\ident) = \ident$].
If it is unital, then Eq.~\eqref{eqt:CP-Crooks} is a fluctuation theorem relating a physical forward and reverse process, where we can interpret $\tilde{P}_{\mE^\ast}(v)$ as the probability density associated with the following \emph{reverse} process (depicted in Fig.~\ref{fig:ReverseProtocol}): i) prepare the state ${\rho}_q := \sum_{\beta} q_\beta \tilde{\rho}_{\beta}$ by measuring $\tilde{\mQ}$, ii) evolve via ${\mE}^{\ast}$, iii) measure $\tilde{\mathcal{ P}}$.  
We emphasize that here, the forward and reverse process are described by \emph{different} measurements, namely ${\mathcal{ P}},{\mathcal{Q}}$ and $\tilde{\mathcal{ P}},\tilde{\mathcal{Q}}$, respectively, related via Eqs.~\eqref{eqt:tildeQ} and \eqref{eqt:tildeP}.

Therefore, Eq.~\eqref{eqt:CP-Crooks} represents a physical fluctuation theorem for unital CPTP maps, \emph{where unitality replaces the role typically occupied by microreversibility}.  Consequently, upon integration of the fluctuation theorem, we obtain $\gamma=1$ (the observation that unital channels yield $\gamma = 1$ was first stated in Ref.~\cite{Kafri:Arxiv}).

Why are unital channels singled out? Recall that the transition matrix $M:=\{p_{\beta|\alpha}\}_{\alpha,\beta}$ of the forward process is in general column stochastic [Eq.~\eqref{eq:colsto}]. Under the additional assumptions of microreversible generalized measurements and unitality, it becomes bistochastic:
\beq
\hspace{-1cm} \sum_\alpha p_{\beta|\alpha} =  \Tr\left[ Q_\beta^\dag Q_\beta {\mE}\left(\sum_\alpha\rho_{\alpha}\right)\right] = \Tr\left[ Q_\beta^\dag Q_\beta {\mE}(\ident)\right] =  \Tr\left[ Q_\beta^\dag Q_\beta \right] = 1\ ,
\label{eq:bisto}
\eeq
Thus, whereas classical microreversibility imposes a specific relation between forward and time-reversed phase space paths \cite{PhysRevE.60.2721}, unitality gives rise to a form of microreversibility relating the forward and reverse probabilities:
\beq
p_{\beta|\alpha} = \Tr \left[ Q_{\beta}^{\dagger} Q_{\beta} \mE \left( \rho_{\alpha} \right) \right] = \Tr \left[ \tilde{P}_{\alpha}^{\dagger} \tilde{P}_{\alpha} \mE^{\ast} \left( \tilde{\rho}_{\beta} \right) \right] = \tilde{p}_{ \alpha|\beta} \ ,
\label{eq:rev}
\eeq 
where Eq.~\eqref{eq:bisto} shows that $\tilde{p}_{ \alpha|\beta}$ is a proper conditional probability.\footnote{
Another unique and suggestive feature of unital channels is that they can always be written as an \emph{affine} combination of reversible channels \cite{Mendl:08}, i.e.,  $\mE(\rho) = \sum_j u_j U_j \rho U_j^\dagger$, where $U_j$ is unitary, $u_j\in \mathbb{R}$ and $\sum_j u_j = 1$.  In the special case when $\{u_j\}$ is a probability distribution, the affine combination becomes a convex one, and the unital channel can be interpreted as representing the unitary evolution $U_j$ occurring with probability $u_j$, while its dual becomes the time-reversed unitary $U_j^\dag$ occurring with the same probability.}

We remark that once we fix the operators $\{P_{\alpha}, Q_{\beta}\}$ in the forward process, then the operators $\{\tilde{P}_{\alpha}, \tilde{Q}_{\beta}\}$ are uniquely defined only for the case of rank-1 projective measurements; otherwise they are defined up to a unitary operator and the full-rank virtual final state $\tilde{\rho}$ [see Eqs.~\eqref{eqt:tildeQ} and \eqref{eqt:tildeP}].  Therefore, beyond rank-1 projective measurements, there is no unique reverse process, but all allowed choices give a fluctuation theorem relating the forward and reverse processes.  

To summarize, the derivation essentially involved only the Kraus representation of CPTP maps, the standard form of generalized measurements, Bayes' rule, and a judicious choice of forward and reverse generalized thermodynamics variables.  The Kraus representation formalism allows for an unambiguous definition of the reverse process in terms of the dual map of the forward process.  However, the choice \eqref{eq:V_choice} is by no means unique.  For example, if we choose $V_{\alpha\beta} = \ln \left( p_{\beta | \alpha} / f_\beta \right)$ (a type of mutual information, as in  \cite{Vedral:2012}) then Eq.~\eqref{eq:3a} yields 
\beq
P_{\mE}(v) e^{-v}= \sum_{\alpha,\beta} \delta(v - V_{\alpha\beta})  p_\alpha f_\beta \ . 
\eeq

%
%%%%%%%%%%%%%%%%%%%%%%%%%%%%%%
\subsection{The Case of Projective measurements} \label{sec:projective}
%%%%%%%%%%%%%%%%%%%%%%%%%%%%%%

When the measurements $\cal P$ and $\cal Q$ are projective, our results for the QJE and fluctuation theorem simplify.  We prepare $\rho_p = \sum_{\alpha}p_\alpha P_\alpha$ and let ${\rho}_q = \sum_\beta q_\beta Q_\beta$, with $\{P_\alpha\}$ and $\{Q_\beta\}$ rank-1 projectors (pure states), and we have $\mathcal{P} = \tilde{\mathcal{P}}$ and $\mathcal{Q} = \tilde{\mathcal{Q}}$, i.e. the forward and reverse processes are described by the same measurements, and Eq.~\eqref{eq:rev} provides a standard microreversibility condition.  Furthermore, in this case, $p_{\alpha}$ and $q_\beta$ can be made arbitrary.  We refer to $\rho_q$ as the ``virtual final state''  since the final state reached at the conclusion of the forward protocol (see Fig.~\ref{fig:Protocol1}) is in general different from the state $\rho_q$.

Using Eq.~\eqref{eq:chidef}, we obtain (more details can be found in \ref{app:B})
\beq
\chi_{\mE}(\lambda) = \Tr \left[ {\rho}_q^{-\lambda } \mE \left( \rho_p^{\lambda+1} \right) \right]\ ,
\label{eq:chi-tr}
\eeq 
and consequently:
\bes
\label{eq:S-proj}
\begin{align}
\label{eqt:Paolo} 
\av{v} &= S({\mE}(\rho_p))- S(\rho_p) +S({\mE}(\rho_p)\|\rho_q) \\
& = S(\rho_q)- S(\rho_p) + \Tr[(\rho_q-{\mE}( \rho_p))\ln(\rho_q) ] \ ,
\label{eqt:Paolo1} 
\end{align}
\ees
where the quantum relative entropy $S\left( \rho \| \sigma \right) = - \Tr [ \rho \ln \sigma ] - S(\rho)$, and $S(\rho) = - \Tr [ \rho \ln \rho ]$ is the von Neumann entropy. This generalizes the result for the mean irreversible entropy production of Refs.~\cite{2010PhRvL.105q0402D,2012PhRvL.109p0601D}.   If $S(\mE(\rho_p)) = S(\rho_p)$ (e.g., when $\mE$ is unitary) then $\av{v} = S\left( \mE(\rho_p) \| {\rho}_q  \right)$.
If $\rho_q$ is $\mE$-invariant [$\mE(\rho_q)  = \rho_q$] and $\mE$ is unitary (as in a quantum quench), then $\av{v} = S\left({\rho}_p \| {\rho}_q \right) $.   If $\mE(\rho_p) = {\rho}_q$ then $\av{v} = S({\rho}_q) - S(\rho_p)$.  If the evolution is adiabatic with initial Hamiltonian $H_i = \sum_{\alpha} \varepsilon_{\alpha}(0)P_{\alpha}$ and final Hamiltonian $H_f = \sum_{\beta} \varepsilon_{\beta}(t_f)Q_{\beta}$ such that $\mE(P_{\alpha}) = Q_{\alpha}$, then $p_{\beta | \alpha} = \delta_{\beta \alpha}$ and  $f_\beta = p_{\beta}$.  Therefore, for the mixed state ensembles $\rho_p = \sum_{\alpha} p_{\alpha} P_{\alpha}$ and $\rho_q = \sum_{\beta} q_{\beta} Q_{\beta}$, we have
\beq
\av{v} = S \left( \mE(\rho_p) \| \rho_q \right) = - \sum_{\alpha} p_{\alpha} \ln q_{\alpha} + \sum_{\alpha} p_{\alpha} \ln p_{\alpha} = H(p \| q)
\eeq
where $(p,q)$ are the probability distributions associated with $\{p_{\alpha} \}$ and $\{q_{\beta} \}$ respectively.

\begin{figure}[t] %  figure placement: here, top, bottom, or page
   \centering
   \includegraphics[width=2.8in]{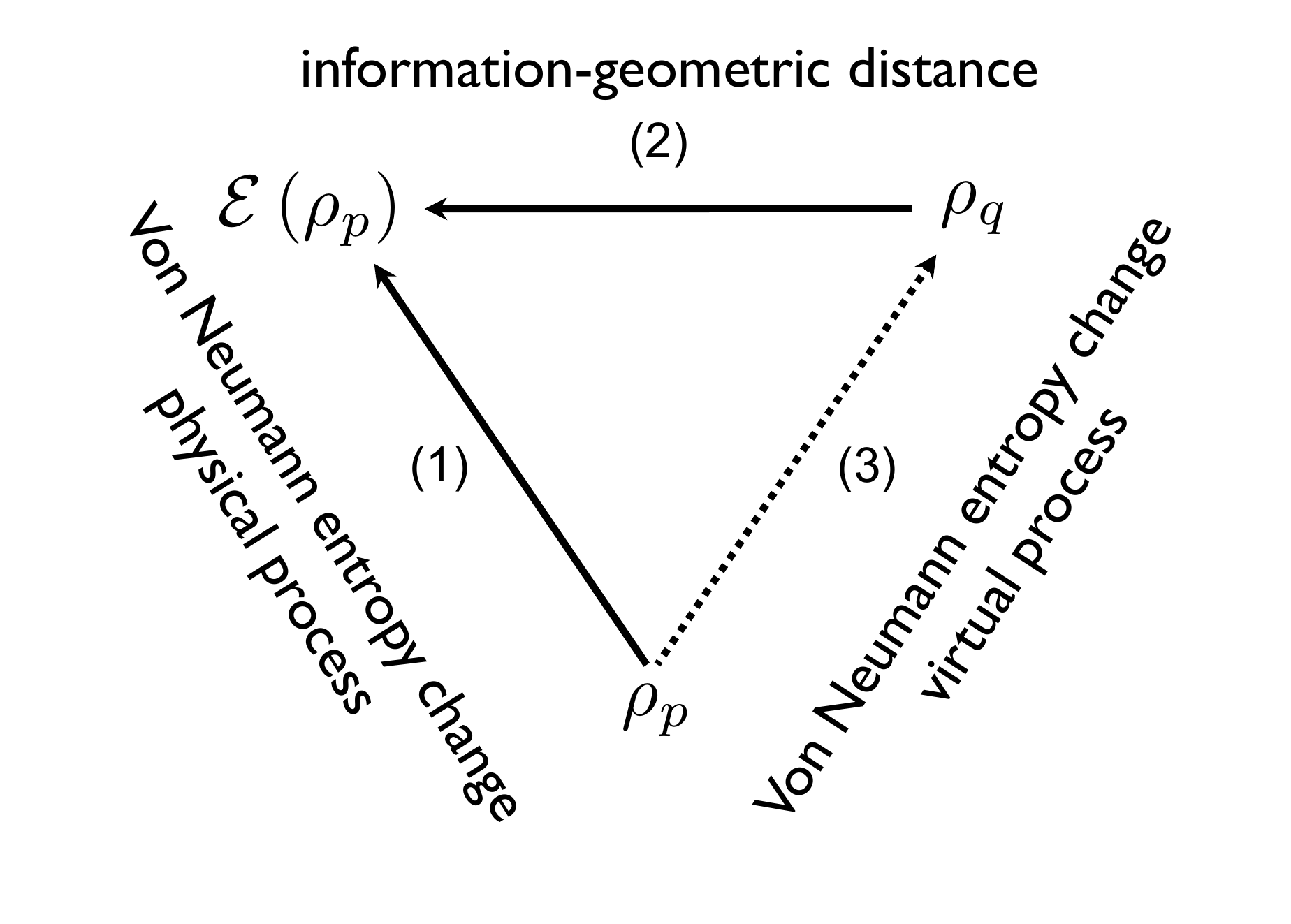} 
   \caption{Distance-entropy diagram illustrating the generalized 2nd law of thermodynamics (see text for details).}   \label{fig:TriangleDiagram}
\end{figure}

Eq.~\eqref{eq:S-proj} has an interesting interpretation, illustrated in Fig.~\ref{fig:TriangleDiagram}. Referring first to Eq.~\eqref{eqt:Paolo}, side $(1)$ of the triangle represents the von Neumann entropy change $(1):=S(\mE(\rho_p))- S(\rho_p)$ occurring in the physical process enacted by $\mE$, while side $(2)$ is an information-geometric measure of the distance $(2):=S(\mE(\rho_p)\|\rho_q)$ between the evolved state ${\mE}(\rho_p)$ and the ``virtual" one $\rho_q$.

On the other hand, referring to Eq.~\eqref{eqt:Paolo1}, side $(2)$ can also represent $(2'):=\Tr[(\rho_q-\mE( \rho_p))\ln (\rho_q)]\le \|{\mE}(\rho_p)-\rho_q\|_1 |\ln (\min_\beta q_{\beta})|$, which is again related to the information-theoretic distance\footnote{Note that the relative entropy is not strictly a distance (since it is not symmetric and does not satisfy the triangle inequality), but a divergence.} between the evolved and virtual state. Side (3) represents the von Neumann entropy change, i.e., $(3):=S(\rho_q)- S(\rho_p)$. 

Known quantum fluctuation theorems follow from our formalism. For example, we show in \ref{app:D} how Eq.~\protect\eqref{eqt:Paolo} reduces to the standard statement of the 2nd law for isothermal processes, $\protect\upbeta \protect\protect\left( \protect\av{w} -  \protect\Delta F \protect\right) = \protect\av{\protect\Delta S_{\mathrm{irr}}} \geq 0$, where $\protect\av{\Delta S_{\mathrm{irr}}}$ is the mean irreversible entropy production \protect\cite{2010PhRvL.105q0402D,2011PhRvL.107n0404D}, after we choose $p$ and $q$ as Gibbs distributions and $\mE$ as a unitary evolution. We also note that a calculation of $\protect\av{v^2} = \left. \frac{d^2 }{d \lambda^2}\chi_{\mE}(\lambda) \right|_{\lambda = 0}$ would yield a generalized fluctuation-dissipation theorem.
 
In the thermal case $\rho_p=e^{-\upbeta H_i}/Z_i$ and $\rho_q=e^{-\upbeta H_f}/Z_f$ (where $H_{i/f}$ is the initial/final Hamiltonian and $Z_{i/f}$ is the initial/final partition function), we find $(2')= \upbeta( \Tr[H_f \mE( \rho_p)]  -\Tr[H_f \rho_q])=:-\upbeta {Q}$. Here $Q$ represents the {\em heat exchange} in the (virtual) undriven relaxation 
between the evolved and virtual states, ${\mE}(\rho_p)\mapsto \rho_q$ (for a proof see \ref{app:A}). In the thermal case $(3)$ amounts to a {\em thermodynamical entropy} change $\Delta S_{qp}$. Thus in this case, using Eq.~\eqref{II} and assuming unitality ($\gamma =1$), we have the 2nd law in the Clausius form $\Delta S_{qp} \geq \upbeta {Q}$. This clarifies why Eq.~\eqref{II} can be interpreted as a generalized 2nd law of thermodynamics.

%
%%%%%%%%%%%%%%%%%%%%%%
\subsection{Including feedback}
\begin{figure}[t] %  figure placement: here, top, bottom, or page
\centering
\includegraphics[width=4in]{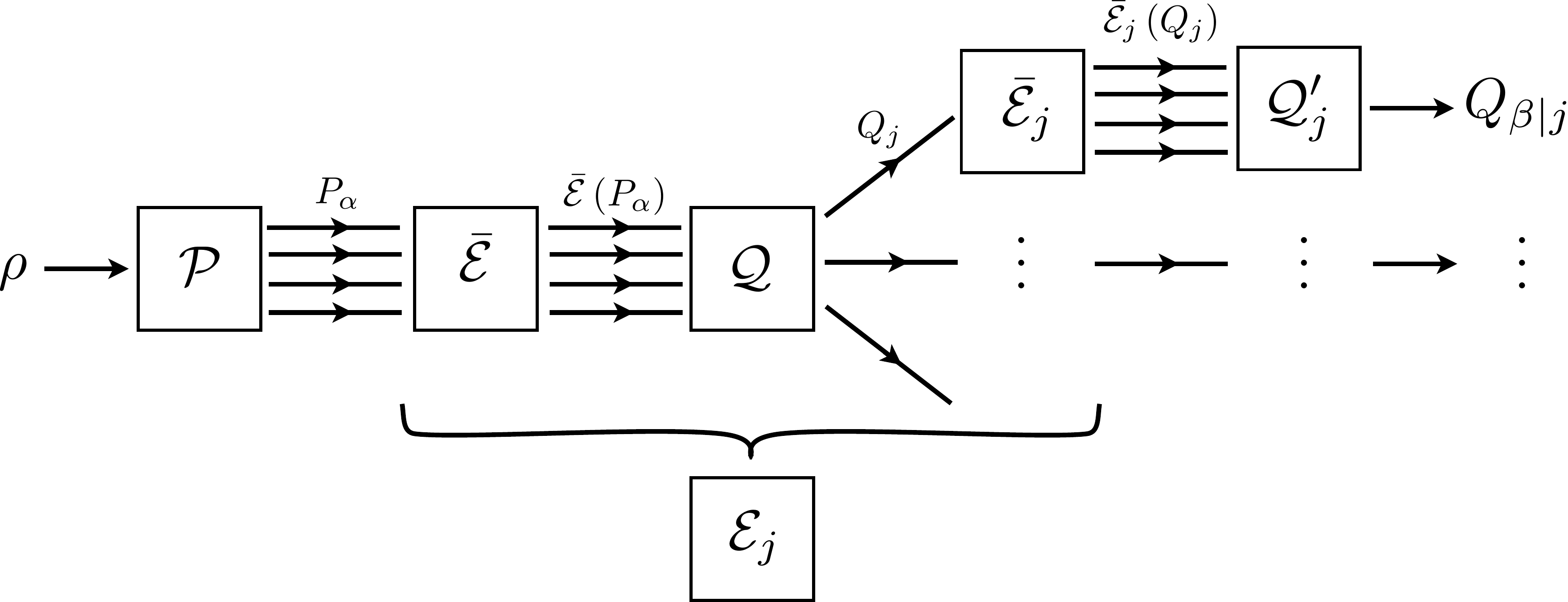} 
\caption{The feedback protocol associated with the forward process with an intermediate measurement whose result introduces the conditional CP map $\mE_j$.}
\label{fig:Protocol2}
\end{figure}

%%%%%%%%%%%%%%%%%%%%%%
%
Suppose we repeat the forward protocol of Fig.~\ref{fig:Protocol1} in the projective measurement case, but denote the CPTP map by $\bar{\mE}$ and the final measurement by $\mQ = \{Q_j\}$. Depending on the measurement outcome $j$, we apply an additional CP map $\bar{\mE}_j$ to the resulting state $Q_j$. This constitutes a \emph{feedback} step, and generalizes earlier work which considered only unitary feedback maps \cite{Sagawa:08,springerlink:10.1007/s10955-011-0153-7,2010PhRvL.104i0602S}. Next we apply another projective measurement $\mQ'_j = \{Q_{\beta|j}\}$, labeled by the outcomes $\beta$. 
Thus $\mE$ from Fig.~\ref{fig:Protocol1} is replaced by $\mE_j := \bar{\mE}_j \circ \mQ \circ \bar{\mE}$, and $\mQ$ by $\mQ'_j$.  Our generalized feedback control protocol is illustrated in Fig.~\ref{fig:Protocol2}.
Given an initial state $P_\alpha$, the probability of observing  outcomes $j$ and $\beta$ is $p_{(j, \beta)| \alpha} = \Tr \left[ Q_{\beta|j} {\mE}_j(P_\alpha) \right]$, and the joint distribution is $p_{(\alpha, j,\beta)} =   p_{(j, \beta)| \alpha} p_\alpha$.
As in the feedback-free case, we can construct the PDF associated with the generalized thermodynamic observable $V$ as:
\bes
\begin{align}
P_{\mE} &:= \sum_{\alpha,j, \beta} \delta (v - V_{\alpha j \beta}) p_{(\alpha, j,\beta)} \\
& = \sum_{\alpha,j, \beta} \delta (v - V_{\alpha j \beta}) \Tr [Q_{\beta|j} {\mE}_j(P_\alpha)] p_\alpha \ , \label{eqt:Feedback_P}
\end{align}
\ees
where the notation $P_{\mE}$ ($\tilde{P}_{\mE^\star}$) is shorthand for $P_{\{{\mE}_j \}}$ ($\tilde{P}_{\{{\mE}^\star_j \}}$).
We now choose $V_{\alpha j \beta} = \ln \left(p_{\alpha}/ q_{\beta|j} \right) = -\tilde{V}_{\beta j \alpha}$, where $q_j:=\{q_{\beta|j}\}$ is an arbitrary, fixed distribution, associated with the (virtual) state $\rho_{q|j} := \sum_{\beta} q_{\beta|j} Q_{\beta|j}$. Then
\bes
\begin{align} 
P_{\mE}(v) e^{-v}  & = \sum_{\alpha, j,\beta} \delta (-v - \tilde{V}_{\beta j \alpha}) \Tr [P_{\alpha} {\mE}^\ast_j(Q_{\beta|j})] q_{\beta|j} \\
& =: \tilde{P}_{\mE^\ast}(-v) \ .
\label{eqt:Feedback_tildeP}
\end{align}
\ees
Integrating, we find a generalized integral fluctuation theorem in the presence of feedback:
\begin{equation}
\langle e^{-v} \rangle = \sum_j \Tr [\rho_{q|j} {\mE}_j(\ident)]= \sum_j \Tr [{\mE}_j^{\ast} \left( \rho_{q|j} \right)] =: \gamma \ . 
\label{eq:gam-feed}
\end{equation}
Generalizing Eqs.~\eqref{eq:chi} and \eqref{eq:chi-tr}, we find (details can be found in \ref{sec:G}) the moment generating function for the feedback case
\beq
\label{eqt:feedback_chi}
\chi_{\mE}(\lambda-1) = \sum_j \Tr \left[ {\rho}_{q|j}^{1-\lambda} {\mE}_j \left( \rho_p^{\lambda} \right) \right]  
%\\ & = \sum_j \Tr \left[  \rho_p^{\lambda}  {\mE}^\ast_j \left( {\rho}_{q|j}^{1-\lambda }\right) \right]  
= \tilde{\chi}_{\mE^\ast}(-\lambda) \ . 
\eeq
We show in \ref{app:E} how these results allow us to recover known quantum fluctuation theorems with feedback.
Although we have identified $P_{\mE}(v) e^{-v} \equiv P_{\{\mE_j\}}(v) e^{-v} $ with $\tilde{P}_{\mE^\ast}(-v) \equiv \tilde{P}_{\{\mE^{\ast}_j\}}(-v)$, it is important to note that $\mE^{\ast}$ does not coincide with the reverse process because (unlike in the feedback-free case) there is now no unique association between the initial $\alpha$ and final $\beta$ measurement outcome indices: the same $(\alpha,\beta)$ pair is connected via different $j$ values of the intermediate measurement outcomes.

%
%%%%%%%%%%%%%%%%%%%%%%
\section{Experimental application of the open-system QJE}
%%%%%%%%%%%%%%%%%%%%%%
Compared to the classical case (e.g., \cite{PhysRevLett.89.050601,Collin:2005,PhysRevLett.102.070602,PhysRevE.71.060101,1742-5468-2005-09-P09011,Ueda:2010,PhysRevB.81.125331,PhysRevX.2.011001,PhysRevLett.109.180601}), there has been relatively little work on experimental tests or applications of the quantum version of the Jarzysnki-Crooks relations.  Existing data on x-ray spectra of simple metals \cite{PhysRevLett.108.190601}, and experiments on a driven single qubit (defect center in diamond)  \cite{PhysRevLett.94.180602} have been used to verify the previously derived quantum fluctuation relations.  There have also been recent proposals that showed that viability of single-qubit interferometry to verify quantum fluctuation theorems \cite{PhysRevLett.110.230601,PhysRevLett.110.230602}.
%, and bidirectional single-electron counting \cite{PhysRevB.81.125331}
Although we are not able to test the generalized QJE or the generalized fluctuation theorem, we present an application of our generalized QJE (specifically, the first moment given by Eq.~\eqref{eq:S-proj}) to the problem of extracting the system-bath coupling magnitude, using both numerical simulations and an experiment using a commercially-available quantum annealing processor comprising superconducting flux qubits (see \ref{app:H} for further details about the experimental system).  Since the system Hamiltonian is time dependent, the formalism of the QJE provides meaningful observables in this setting.

The processor performs a quantum annealing protocol to find the ground state of a classical Ising Hamiltonian.  The protocol is described by the transverse-field Ising Hamiltonian 
\beq \label{eqt:QA_H}
H_S(t) = -A(t) \sum_i \sigma^x_i + B(t) H_{\mathrm{Ising}} \ , 
\eeq
where 
\beq
H_{\mathrm{Ising}} = -\sum_{i=1}^N h_i \sigma^z_i - \sum_{i<j}^N J_{ij} \sigma^z_i \sigma^z_j \ ,
\eeq
and $\sigma_i^{x,z}$ are standard Pauli operators acting on the $i$th qubit. The magnetic fields $h_i$ and couplings (superconducting inductances) $J_{ij}$ are programmable.  The annealing functions $A(t)$ and $B(t)$ satisfy $A(0),B(t_f) \neq 0, \ A(t_f), B(0) = 0$, where $t_f$ is the total annealing time. The annealing protocol amounts to starting with the transverse field turned on and the Ising Hamiltonian turned off, and then slowly reversing their role until only the Ising Hamiltonian remains. The processor is performed at $T=17$mK, with the qubits in contact with a thermal environment.  It is ideally suited to measuring $\av{v}$ in Eq.~\eqref{eq:S-proj} since it performs the process described in Fig.~\ref{fig:Protocol1} (but without the measurement $\mathcal{P}$) with the initial state being the Gibbs state.  

The experiment can be described by an adiabatic Markovian master equation derived in Ref.~\cite{2012arXiv1206.4197A} (see \ref{app:F} for essential details).  The CPTP map $\mE$ generated by the master equation is \emph{not} unital.  We consider projective measurements that prepare $p_\alpha = e^{-\upbeta  \varepsilon_\alpha(0)}/ Z(0)$, $q_\beta =  e^{-\upbeta  \varepsilon_\beta(t_f) }/Z(t_f)$, where $\varepsilon_{\alpha,\beta}(t)$ are the instantaneous eigenenergies of $H_S(t)$, so that our generalized thermodynamic observable [Eq.~\eqref{eq:V_choice}] is given by $V_{\alpha\beta} = \upbeta( \varepsilon_\beta(t_f) - \varepsilon_\alpha(0) ) -\upbeta(F(t_f) - F(0))$, where the free energy $F = -\ln( Z)/\upbeta$.  Note that for an open quantum system, $\varepsilon_\beta(t_f) -\varepsilon_\alpha(0)$ does \emph{not} correspond to the work done on the system.  Equations \eqref{eq:gamma} and \eqref{eqt:albash} yield
\begin{equation} 
\label{eqt:Specific}
\langle e^{-\upbeta ( \Delta E - \Delta F)} \rangle = \gamma = \Tr \left[ \mE^{\ast}(\rho_{\mG}(t_f)) \right] \ ,
\end{equation}
where $\rho_{\mG}(t) = \exp(-\upbeta H_S(t))/Z(t)$ denotes the Gibbs state associated with $H_S(t)$, with $\rho_p \equiv \rho_{\mG}(0)$ and $\rho_q \equiv \rho_{\mG}(t_f)$, $\Delta E $ is a random variable taking values in the set $\{\varepsilon_\beta(t_f) - \varepsilon_\alpha(0)\}$, and $\Delta F:=F(t_f) - F(0)$.  Equation~\eqref{eqt:Paolo} gives:
\begin{eqnarray}
\upbeta \left(\langle \Delta E \rangle - \Delta F \right) &=& S \left(\rho(t_f) \| \rho_{\mG}(t_f) \right) + S(\rho(t_f)) - S \left( \rho_{\mG}(0) \right) \ ,
\end{eqnarray}
where we have denoted $\rho(t_f) = \mE(\rho_{\mG}(0))$.

\begin{figure}[t] %  figure placement: here, top, bottom, or page
\centering
\includegraphics[width=2.8in]{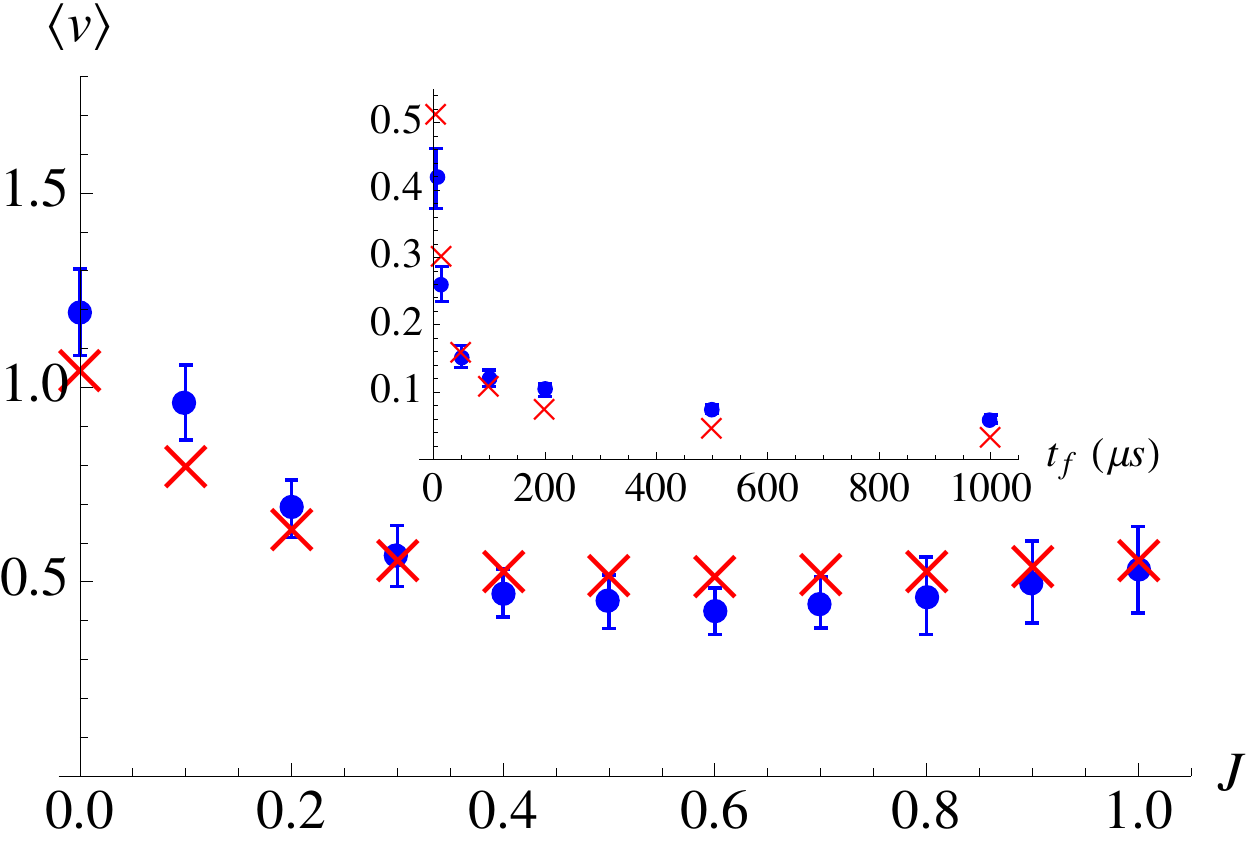}
\caption{Experimental results (blue dots with error bars) for $\av{v}$ [from Eq.~\eqref{eq:<v>}] and the best fit using the adiabatic Markovian master equation (red $\times$) with the extracted value $\kappa =2.34 \times 10^{-3}$.  Main panel: as a function of $J$, with $t_f=5\mu$s in both the numerical simulations and the experiment. Inset: as a function of $t_f$, with $J = 1/2$ in both the numerical simulations and the experiment.}
%
%\caption{Experimental results (blue dots with error bars) for $\av{v}$ [from Eq.~\eqref{eq:<v>}] and the best fit using the adiabatic Markovian master equation (red $\times$) with the high-frequency cut-off $\omega_c = 8 \pi$ and the extracted value $\kappa:=g^2 \eta/\hbar^2 = 5.75/\pi \times 10^{-3}$, where $g$ is the  system-bath coupling constant and $\eta$ characterizes the Ohmic bath \cite{sup-mat}.  Main panel: as a function of ferromagnetic coupling $J$, with $t_f=5\mu$s in both the numerical simulations and the experiment. Inset: as a function of total annealing time $t_f$, with $J = 1/2$ in both the numerical simulations and the experiment.}
\label{fig:AverageW_Experiment}
\end{figure}

For concreteness we consider a two-qubit system with $H_{\mathrm{Ising}} = - \frac{1}{3} \sum_{i=1}^2 \sigma^z_i - J \sigma^z_1 \sigma^z_2$.
We checked the equality expressed in Eq.~\eqref{eqt:Specific} by independently numerically simulating its two sides for this model using our master equation, and find essentially perfect agreement. The same holds for Eq.~\eqref{eqt:Paolo} (see \ref{app:G} for details). 
We tested the same two-qubit system experimentally using the quantum annealing processor.  For each experimental run, the system is initialized in the Gibbs state of $H_S(0)$, and after performing the annealing protocol $H_S(t)$, a projection onto the computational basis (eigenstates of the $\sigma_i^z$ operators, i.e., the final energy eigenstate basis of $H_{\mathrm{Ising}}$) is performed.  Therefore, for each run, a single energy eigenstate of $H_{\mathrm{Ising}}$ is measured.  We then repeat the quantum annealing process thousands of times to build up the statistics necessary to determine the relative occupancy of each final energy eigenstate. The empirical relative occupancy corresponds directly to the probability $f_\beta$ of measuring energy $\varepsilon_\beta(t_f)$.  
Therefore, this allows us to experimentally determine $\av{\varepsilon (t_f)} = \sum_\beta \varepsilon_\beta (t_f) f_\beta$. We also know $\av{\varepsilon(0)}$ from the initial Gibbs state and the value of $A(0)$. We thus determine
\bes
\begin{align}
\av{v} &= \upbeta \sum_{\alpha,\beta}  \left(\varepsilon_\beta(t_f) - \varepsilon_\alpha(0) - \Delta F \right) p_{(\alpha,\beta)} \nonumber \\
%&= \upbeta \left( \sum_\beta \varepsilon_\beta(t_f) p_\beta - \sum_\alpha \varepsilon_\alpha(0) p_\alpha - \Delta F \right) \nonumber \\
&= \upbeta \left( \av{\varepsilon(t_f)}  - \av{\varepsilon(0)} - \Delta F \right)\ ,
\label{eq:<v>}
\end{align}
\ees
where we compute $\Delta F$ using exact diagonalization (see \ref{app:G} for details). We show these results in Fig.~\ref{fig:AverageW_Experiment} 
as a function of $t_f$ and ferromagnetic coupling strength $J$.
Our master equation has two free parameters: the high-frequency cut-off $\omega_c$, which we set to  $8 \pi$ \cite{2012arXiv1206.4197A}, and the system-bath coupling magnitude $\kappa=g^2 \eta/\hbar^2$, where $g$ is the  system-bath coupling constant and $\eta$ characterizes the Ohmic bath (see \ref{app:F} for details). $\langle v \rangle$ is a quantity that combines both the statistics and the energy spectrum of the system, making it more system-specific.  Remarkably, by simultaneously minimizing the deviation between the numerical solution of our master equation and the experimental data for $\av{v}$ as a function of $J$ and $t_f$ allowed us to extract the system-bath coupling magnitude $\kappa$ from Eq.~\eqref{eq:<v>} (see \ref{app:H} for details).  Therefore, we find that $\langle v \rangle$ provides a valuable optimization target in addition to its information theoretic content, which the statistics of the experiment alone may not provide.  

Why does $\av{v}$ display a minimum as a function of $J$  (main panel of Fig.~\ref{fig:AverageW_Experiment})? In our experiment $\upbeta A(0) \sim \upbeta B(t_f) \approx 15$, so that the  Gibbs state is almost pure, i.e., both $S(\rho_\mG(0))$ and $S(\rho_\mG(t_f)) \ll 1$. Therefore \emph{we are effectively measuring the information-theoretic distance} $S \left(\rho(t_f) \| \rho_{\mG}(t_f) \right)$. Increasing $J$ at fixed temperature $kT=1/\upbeta$ is like decreasing $T$ while fixing $J$. Thus the system requires more time to equilibrate as $J$ grows, but we keep $t_f$ fixed. On the other hand, as $J$ becomes very small the ground and first excited states become degenerate, so the excitation probability increases, and the system is again farther from equilibrium. Also, as we increase the annealing time, $\rho(t_f)$ becomes closer to the Gibbs state, causing $S \left(\rho(t_f) \| \rho_{\mG}(t_f) \right)$ to decrease as observed in the inset.  

\section{Conclusions}
To conclude, we presented fluctuation theorems and moment generating functions for CPTP maps, thus generalizing previous work on the Jarzynski-Crooks relations and the 2nd law for open quantum systems, including processes with feedback. 
We performed an experiment using superconducting flux qubits that matches the fluctuation theorem protocols, and used this experiment to extract the system-bath coupling for an adiabatic Markovian master equation that nicely matches the experimental results. Our work 
ties together key ideas from statistical mechanics, quantum information theory, and the theory of open quantum systems, and 
paves the way to experimental tests and applications of fluctuation theorems in the most general setting of open quantum system dynamics.

Note added: After the appearance of our work on the arXiv, two papers arrived at similar results \cite{2013JSMTE..06..016R,2013arXiv1307.5370R}.

\acknowledgments
We thank S. Deffner, P. Talkner, and I. Marvian for useful comments.  This research was supported by the ARO MURI grant W911NF-11-1-0268, the Lockheed Martin Corporation, and by NSF grant numbers PHY- 969969 and PHY-803304 (to P.Z. and D.A.L.). 

\appendix
%%%%%%%%%%%%%%%%%%%%%%%%%%%%%%%%%%%%%%%%%%%%

%%%%%%%%%%%%%%%%%%%%%%
\section{Recovering Known Fluctuation Theorems} \label{app:D}
%%%%%%%%%%%%%%%%%%%%%%
%
In order to recover the well-established closed system results \cite{2000cond.mat..9244T}, we consider the CPTP map $\mE$ to be simply the unitary evolution, as in Section~\ref{sec:closed-QJE}:
\begin{equation}
\mE(X) = U(t_f,0) X U^{\dagger}(t_f,0) \ .
\end{equation}
Since this map is unital, its dual is also a CPTP map (the actual time reversed process) given by:
\begin{equation}
\mE^{\ast}(X) = U^\dagger(t_f,0) X U(t_f,0) \ ,
\end{equation}
where $U(t_f,0) = T_+ \exp(-i\int_0^{t_f} H(t)dt)$, and the Hamiltonian $H(t)$ has instantaneous eigenenergies $\varepsilon(t)$. %Using Eq.M.~(6) this yields $\gamma = 1$.  
Using Eq.~\eqref{eq:gamma} this yields $\gamma = 1$.  

Recall that in our formalism we need to also specify the fiducial initial state $\rho$, the measurements $\mP$ and $\mQ$, and the distribution $q$. We pick these so that they generate the Gibbs distributions
\begin{equation}
p_\alpha = \frac{e^{-\upbeta  \varepsilon_\alpha(0) }}{Z(0)} \ , \quad q_\beta =  \frac{e^{-\upbeta  \varepsilon_\beta(t_f) } }{Z(t_f)}
\end{equation}
at inverse temperature $\upbeta$, where $Z(t) = \Tr[\exp(-\upbeta H(t)]$ is the partition function corresponding to $H(t)$, and $\rho_{\mG}(t) = \exp(-\upbeta H(t))/Z(t)$ is the corresponding Gibbs state. For example, we can assume that $\rho = \rho_{\mG}(0)$, $\mP = \{\ket{\varepsilon_\alpha(0)}\bra{\varepsilon_\alpha(0)}$, and $\mQ = \{\ket{\varepsilon_\beta(t_f)}\bra{\varepsilon_\beta(t_f)}\}$.

If we let $V_{\{\alpha, \beta\}}=\ln(p_\alpha/q_\beta)$ this then corresponds to the choice:
\begin{equation}
V_{\{\alpha, \beta\}} = \upbeta( \varepsilon_\beta(t_f) - \varepsilon_\alpha(0) ) - \upbeta(F(t_f) - F(0)) \ , 
\end{equation}
where $F$ is the free energy given by $ F(t) = - \ln Z(t) / \upbeta$. This corresponds to identifying $v$ with the (dimensionless) work.
  
%We thus find from Eq.M.~(7) the QJE for a closed quantum system:
We thus find from Eq.~\eqref{eqt:albash} the QJE for a closed quantum system:
\begin{equation}
\langle e^{-\upbeta ( \Delta E - \Delta F)} \rangle = 1\ ,
\end{equation}
where $\Delta E $ is a random variable taking values in the set $\{\varepsilon_\beta(t_f) - \varepsilon_\alpha(0)\}$, and $\Delta F:=F(t_f) - F(0)$. For these choices, Eq.~\eqref{eq:S-proj} becomes
\begin{eqnarray}
\beta \left( \langle \Delta E \rangle -  \Delta F \right) &=& S \left(\mE(\rho_{\mG}(0)) \| \rho_{\mG}(t_f) \right)   \ ,
\label{eq:34}
\end{eqnarray}
where we have used $ S(\mE(\rho_{\mG}(0))=S \left( \rho_{\mG}(0) \right)$.  For a closed quantum system, the heat transfer in and out of the system is zero, so $\Delta E$ is equal to the work $w$.  Furthermore, for a closed quantum system, the relative entropy $S \left(\mE(\rho_{\mG}(0)) \| \rho_{\mG}(t_f) \right)$ is equal to the mean irreversible entropy production $\av{\Delta S_{\mathrm{irr}}}$ \cite{2010PhRvL.105q0402D,2011PhRvL.107n0404D}, so we can rewrite Eq.~\eqref{eq:34} as
\begin{equation}
\upbeta \left( \av{w} -  \Delta F \right) = \av{\Delta S_{\mathrm{irr}}}  \ ,
\end{equation}
which is the 2nd law of thermodynamics, since $\av{\Delta S_{\mathrm{irr}}}\geq 0$. 

As this example illustrates, our formalism clarifies the subtle relationship between the choice of the thermodynamic observable $V_{\{\alpha, \beta\}}$, the initial state probability $p_\alpha$, and the dual reverse state probability $q_\beta$, which together comprise the Jarzynski equality and lead to the 2nd law.

\section{Generalized measurements that restrict the probability $p_\alpha$} \label{app:A1}
%%%%%%%%%%%%%%%%%%%%%%%%%%%%%%%%%%%%%%%%%%%%
Let us consider the generalized measurements:
\beq
P_1 = \left( \begin{array}{cc}
\frac{1}{\sqrt{3}} & 0 \\
0 & \sqrt{\frac{2}{3}} 
\end{array} \right) \ , \quad P_2 = \left( \begin{array}{cc}
\sqrt{\frac{2}{3}} & 0 \\
0 & \frac{1}{\sqrt{3}} 
\end{array} \right) \ .
\eeq
Let us consider applying the measurements on the arbitrary state:
\beq
\rho = \left( \begin{array}{cc}
a & b \\
b^\ast & 1-a
\end{array} \right) \ ,
\eeq
where $0 \leq a \leq 1$ and $b$ is a complex number.  The possible resulting states are:
\beq
\rho_1 =  \frac{3}{2-a} \left( \begin{array}{cc}
\frac{a}{3} & b \\
b^\ast & \frac{2}{3}(1-a) 
\end{array} \right) \ , \quad \rho_2 =  \frac{3}{1+a} \left( \begin{array}{cc}
\frac{2}{3}a & b \\
b^\ast & \frac{1}{3}(1-a) 
\end{array} \right) 
\eeq
with probabilities $p_1 = (2-a)/3$ and $p_2 = (1+a)/3$ respectively.  Therefore we find that $ 1/3 \leq p_1 \leq 2/3$ and $ 1/3 \leq p_2 \leq 2/3$, so we cannot make them take arbitrary values by an appropriate choice of $\rho$.
%
%%%%%%%%%%%%%%%%%%%%%%%%%%%%%%%%%%%%%%%%%%%%
\section{Derivation of the entropy formulas for $\av{v}$} \label{app:A}
%%%%%%%%%%%%%%%%%%%%%%%%%%%%%%%%%%%%%%%%%%%%
%
%\subsection{Generalized measurements: Eq.M.~(11)}
\subsection{Generalized measurements: Eq.\eqref{eq:ent-gen-meas}}

%We prove Eq.M.~(11) 
We prove Eq.\eqref{eq:ent-gen-meas} from the first moment expression $\av{v} = \frac{d}{d \lambda}\chi_{\mE}(\lambda) \Big|_{\lambda = 0}$. Then, using Eq.~\eqref{eq:18a}

\bes
\begin{align}
\av{v} &= \frac{d}{d \lambda} \Tr\left[ \mE \left(\sum_\alpha p_\alpha ^{\lambda+1}\rho_\alpha\right)\sum_\beta q_\beta^{-\lambda} \tilde{\rho}_{\beta}\right]\Big|_{\lambda = 0} \\
& = \Tr\left[ \mE \left(\sum_\alpha p_\alpha ^{\lambda+1} \ln(p_\alpha) \rho_\alpha\right)\sum_\beta q_\beta^{-\lambda} \tilde{\rho}_{\beta} \right]\Big|_{\lambda = 0}  \\
& \quad - \Tr\left[
\mE \left(\sum_\alpha p_\alpha ^{\lambda+1}\rho_\alpha\right)\sum_\beta q_\beta^{-\lambda} \ln(q_\beta) \tilde{\rho}_{\beta}
\right]\Big|_{\lambda = 0} \\
\label{eq:21d}
& = \Tr\left[ \mE\left(\sum_\alpha p_\alpha \ln(p_\alpha) \rho_\alpha\right)\right]-\Tr\left[\mE(\rho_p)\sum_\beta\ln(q_\beta)\tilde{\rho}_{\beta}\right] \\
\label{eq:21e}
& = \sum_\alpha p_\alpha \ln(p_\alpha) - \sum_\beta f_\beta \ln(q_\beta) \ ,
\end{align}
\ees
where to arrive at Eq.~\eqref{eq:21d} we used the fact that $\sum_\beta\tilde{\rho}_{\beta} = \sum_\beta Q_\beta^\dag Q_\beta = \ident$, to arrive at Eq.~\eqref{eq:21e} that $\mE$ is trace-preserving and $\Tr[\rho_\alpha]=1$, and observed that 
$\Tr\left[\tilde{\rho}_{\beta} \mE(\rho_p)\right] = \Tr\left[Q_\beta^\dag Q_\beta \mE(\sum_{\alpha} p_\alpha \rho_\alpha)\right]=\sum_{\alpha} \Tr[Q_\beta^\dag Q_\beta \mE(\rho_\alpha)]p_\alpha= \sum_{\alpha} p_{\beta|\alpha}p_\alpha  = f_\beta $. 
Adding and subtracting $H(f) = -\sum_\beta f_\beta \ln(f_\beta)$ we thus arrive at Eq.~\eqref{eq:ent-gen-meas}.

%\subsection{Projective measurements:  Eq.M.~(13)}
\subsection{Projective measurements:  Eq.~\eqref{eq:S-proj}}
%
%Next we prove Eq.M.~(13a) using a similar technique. Starting from Eq.M.~(12) we have
Next we prove Eq.~\eqref{eqt:Paolo} using a similar technique. Starting from Eq.~\eqref{eq:chi-tr} we have
\bes
\begin{align}
\av{v} & =  \frac{d}{d \lambda}\Tr \left[ {\rho}_q^{-\lambda } \mE \left( \rho_p^{\lambda+1} \right) \right] \Big|_{\lambda = 0} \\
& = \Tr \left[{\rho}_q^{-\lambda } \mE \left( \rho_p^{\lambda+1} \ln(\rho_p)\right) -{\rho}_q^{-\lambda } \ln(\rho_q)\mE \left( \rho_p^{\lambda+1} \right) \right] \Big|_{\lambda = 0} \\
\label{eq:24c}
& =  \Tr \left[ \mE \left( \rho_p \ln(\rho_p)\right)\right] -\Tr \left[\ln(\rho_q)\mE \left( \rho_p \right) \right] \\
%& = \Tr \left[ \rho_p \ln(\rho_p) \right]-\Tr \left[\ln(\rho_q)\mE \left( \rho_p \right) \right]
% +S \left( \mE \left( \rho_p \right)\right)-S \left( \mE \left( \rho_p \right)\right)\\
\label{eq:24e}
& = S\left( \mE(\rho_p) \| {\rho}_q \right) + S \left( \mE \left( \rho_p \right) \right) - S \left( \rho_p \right)\ ,
\end{align}
\ees
where to arrive at Eq.~\eqref{eq:24e} we used the fact that $\mE$ is trace-preserving, and added and subtracted $S \left( \mE \left( \rho_p \right)\right)$.

%Finally, Eq.M.~(13b) amounts to the following calculation, starting from Eq.~\eqref{eq:24c}:
Finally, Eq.~\eqref{eqt:Paolo1} amounts to the following calculation, starting from Eq.~\eqref{eq:24c}:
\bes
\begin{align}
& \Tr \left[ \mE \left( \rho_p \ln(\rho_p)\right)\right] -\Tr \left[\ln(\rho_q)\mE \left( \rho_p \right) \right] \\
&\quad = S(\rho_q) + \Tr \left[ \rho_p \ln(\rho_p)\right] -\Tr \left[\ln(\rho_q)\mE \left( \rho_p \right) \right] - S(\rho_q)\\
&\quad = S(\rho_q)- S(\rho_p) + \Tr[(\rho_q-{\mE}( \rho_p))\ln \rho_q ] \ .
\end{align}
\ees

\subsection{The heat term}
In the main text we claimed that in the thermal case, when $\rho_q=e^{-\upbeta H_f}/Z_f$ (where $H_{f}$ is the final Hamiltonian), we find $(2'):=\Tr[(\rho_q-\mE( \rho_p))\ln (\rho_q)]= -\upbeta Q$. Here is the proof:
\bes
\begin{align}
\label{eq:27a} \Tr[(\rho_q-\mE( \rho_p))\ln (\rho_q)] & =  -\Tr[(\rho_q-\mE( \rho_p))(\upbeta H_f +\ln Z_f)]  \\
&= \upbeta( \Tr[H_f \mE( \rho_p)]  -\Tr[H_f \rho_q]) \\
& = \ \upbeta (\av{H_f}_{\mE( \rho_p)} - \av{H_f}_{\rho_q}) \ ,
\label{eq:27d}
\end{align}
\ees
where to arrive at the second equality of Eq.~\eqref{eq:27a} we used $\Tr[\mE( \rho_p)\ln Z_f] = \ln Z_f$ since $\mE$ is trace-preserving, and $\Tr[\rho_q \ln Z_f] =  \ln Z_f$. Since the (virtual) relaxation process from the state $\mE( \rho_p)$  to the state $\rho_q$ is undriven (i.e., there is no work involved), the internal energy change expressed in Eq.~\eqref{eq:27d} is a pure heat exchange.

%%%%%%%%%%%%%%%%%%%%%%%%%%%%%%%%%%%%%%%%%%%%
\section{Derivation of the number of elements in Eq.~\eqref{eqt:constraint}} \label{app:A2}
%%%%%%%%%%%%%%%%%%%%%%%%%%%%%%%%%%%%%%%%%%%%
%
We prove that the constraints in Eq.~\eqref{eqt:constraint} require that the number of measurement operators in $\mathcal{P}$ and $\mathcal{Q}$ (denoted $N_p$ and $N_q$ respectively) must equal the dimension of the Hilbert space $d$.  First, consider the constraint on  $\mathcal{P}$ and consider $\rho$ to be the maximally mixed state:
\beq
\sum_{\alpha=1}^{N_p} \rho_\alpha = \sum_{\alpha=1}^{N_p} \frac{P_{\alpha} \ident P_{\alpha}^{\dagger}}{\Tr \left( P_{\alpha}^{\dagger} P_{\alpha} \right) } = \ident \ .
\eeq
Taking the trace on both sides yields $N_p = d$.  Similarly, consider the constraint $\sum_\beta Q_{\beta}^{\dagger} Q_{\beta}= \ident$ on  $\mathcal{Q}$  and again take the trace:
\beq 
\sum_{\beta = 1}^{N_q} \Tr \left( Q_{\beta}^{\dagger} Q_{\beta} \right) = \sum_{\beta = 1}^{N_q} 1 =d \ .
\eeq
Thus $N_q = d$.

%%%%%%%%%%%%%%%%%%%%%%%
\section{Bounding the value of $\gamma$ for microreversible generalized measurements}
%%%%%%%%%%%%%%%%%%%%%%%%
%
\label{app:L}

Let us denote the trace-norm by $|| \cdot ||_1$.  It is defined by:
\beq
|| A ||_1 \equiv \mathrm{Tr} | A | = \sum_i s_i(A) \ ,
\eeq
where $|A| = \sqrt{A^\dagger A} $ and $s_i(A)$ are the singular values of $A$.  Let us also define supoperator norm $|| \cdot ||_{\infty}$ defined by
\beq
|| A ||_{\infty} = \sup_{| \psi \rangle : \langle \psi | \psi \rangle = 1} \sqrt{\langle \psi | A^\dagger A | \psi \rangle } = \max_{i} s_i(A) \ .
\eeq
Consider two hermitian operators $A = A^{\dagger}$ and $B = B^{\dagger}$ with $||A||_1 = ||B||_1 = 1$.  Let us consider the quantity $c := || A B ||_1 = \mathrm{Tr} |A B |$.  It satisfies:
\beq
c \leq || A ||_1 ||B||_{\infty} = ||B||_{\infty}  \ , \quad c \leq || B ||_1 || A ||_{\infty} = || A ||_{\infty}  \ .
\eeq
(For more details about norms and inequalities between them see, e.g., Refs.~\cite{Bhatia:1997:SpringerVerlag,Watrous:2004:lecture-notes,Lidar:2008:012308}.)  Then $c$ must satisfy
\beq
c \leq \min \left\{ || A ||_{\infty} , || B ||_{\infty} \right\} \ .
\eeq
Therefore for $A = \rho_q$ and $B = \mathcal{E}(\ident /d)$, where $d$ is the dimension of the Hilbert space, we have an upper bound for $\gamma$ given by
\beq
0 \leq \gamma = \mathrm{Tr} \left[ \rho_q \mathcal{E}(\ident) \right] \leq d \min \left\{ || \rho_q ||_{\infty} , || \mathcal{E} \left(\ident/d \right) ||_{\infty} \right\} \leq d \ .
\eeq
If either $\rho_q$ or $\mathcal{E} \left(\ident/d \right)$ is maximally mixed then the bound becomes $\gamma \leq 1$, which is as tight as possible.  An example that shows, depending on the choice of $\rho_q,$ that $\gamma$ can take all possible values in $[0,\,d]$ is provided by the amplitude damping channel where
\beq
\mE(\rho) = \ketbra{1}{1} \ ,
\eeq
for all states $\rho$, which gives $\gamma = d \langle 1 | \rho_q | 1 \rangle$.

%%%%%%%%%%%%%%%%%%%%%%%%%%%%%%%%%
%\section{Derivation of Eq.M.~(10) and Eq.M.~(12)}
\section{Derivation Eq.~\eqref{eq:chi-tr}} \label{app:B}
%Using Eq.M.~(3b) and Eq.M.~(9) we have
Using Eq.~\eqref{eq:3b} and Eq.~\eqref{eq:chidef} we have
\bes
\begin{align}
\chi_{\mE}(\lambda) &:= \int_{-\infty}^{\infty} d v P_{\mE}(v) e^{\lambda v}  \\
&= \int_{-\infty}^{\infty} d v e^{\lambda v}\sum_{\alpha, \beta} \delta(v -V_{\{\alpha, \beta\}}) \Tr[\tilde{\rho}_{\beta} {\mE}(\rho_{\alpha})] p_\alpha \\
& = \sum_{\alpha,\beta} \left({p_\alpha}/{q_\beta}\right)^\lambda \Tr\left[\tilde{\rho}_{\beta} \mE(\rho_\alpha)\right]p_\alpha \\
\label{eq:18a}
& = \Tr\left[ \mE \left(\sum_\alpha p_\alpha ^{\lambda+1}\rho_\alpha\right)\sum_\beta q_\beta^{-\lambda} \tilde{\rho}_{\beta}\right] ,
\end{align}
\ees
so that
\bes
\begin{align}
\chi_{\mE}(\lambda-1)  &=  \Tr\left[ \mE \left(\sum_\alpha p_\alpha ^{\lambda}\rho_\alpha\right)\sum_\beta q_\beta^{1-\lambda} \tilde{\rho}_{\beta}\right] \\
& = \Tr\left[ \sum_\alpha p_\alpha ^{\lambda}\rho_\alpha \, \mE^* \left(\sum_\beta q_\beta^{1-\lambda} \tilde{\rho}_{\beta}\right)\right] \  .
\end{align}
\ees
%On the other hand, using Eq.M.~(4b) and Eq.M.~(9) we have
On the other hand, using Eq.~\eqref{eq:4b} and Eq.~\eqref{eq:chidef} we have
\bes
\begin{align}
\tilde{\chi}_{\mE^*}(-\lambda) &= \int_{-\infty}^{\infty} d v \tilde{P}_{\mE^*}(v) e^{-\lambda v}  \\
&=  \int_{-\infty}^{\infty} d v e^{-\lambda v}\sum_{\alpha, \beta} \delta(v -\tilde{V}_{\{\beta, \alpha\}}) \Tr[{\rho}_\alpha {\mE^*}(\tilde{\rho}_{\beta})] q_\beta \\
&= \sum_{\alpha,\beta} \left({p_\alpha}/{q_\beta}\right)^\lambda \Tr\left[\rho_\alpha \mE^*(\tilde{\rho}_{\beta})\right]q_\beta  \\
&= \Tr\left[ \sum_\alpha p_\alpha ^{\lambda}\rho_\alpha\, \mE^* \left(\sum_\beta q_\beta^{1-\lambda} \tilde{\rho}_{\beta}\right) \right]  = \chi_{\mE}(\lambda-1) \ .
\end{align}
\ees

In the case of projective measurements and and when $\rho_p = \sum_{\alpha}p_\alpha P_\alpha$, ${\rho}_q = \sum_\beta q_\beta Q_\beta$, with $\{P_\alpha\}$ and $\{Q_\beta\}$ rank-1 projectors, Eq.~\eqref{eq:18a} %directly becomes Eq.M.~(12).
directly becomes Eq.~\eqref{eq:chi-tr}.

%%%%%%%%%%%%%%%%%%%%%%
%\section{Derivation of Eq.M.~(17)}
\section{Derivation of Eq.~\eqref{eqt:feedback_chi}} \label{sec:G}
%%%%%%%%%%%%%%%%%%%%%%
%
%Using Eq.M.~(9) and Eq.M.~(14b), we have
Using Eq.~\eqref{eq:chidef} and Eq.~\eqref{eqt:Feedback_P}, we have
\bes
\begin{align}
\chi_{\mE} (\lambda)&:= \int_{-\infty}^{\infty} d v P_{\mE}(v) e^{\lambda v} \\
&= \int_{-\infty}^{\infty} d v e^{\lambda v} \sum_{\alpha, j, \beta} \delta(v - V_{\alpha j \beta}) \Tr \left[ Q_{\beta|j} {\mE}_j(P_{\alpha}) \right] p_{\alpha} \\
&= \sum_{\alpha, j, \beta}(p_{\alpha}/q_{\beta|j})^{\lambda}  \Tr \left[ Q_{\beta|j} {\mE}_j(P_{\alpha}) \right] p_{\alpha}  \\
&=  \sum_j\Tr \left[  {\mE}_j \left( \sum_{\alpha} p_{\alpha}^{\lambda+1} P_{\alpha} \right) \sum_{\beta} q_{\beta|j}^{-\lambda} Q_{\beta|j} \right] \\
&= \sum_j \Tr \left[  {\mE}_j \left( \rho_p^{\lambda+1} \right) \rho_{q|j}^{-\lambda} \right] \ ,
\end{align}
\ees
so that
\beq
\chi_{\mE} (\lambda-1)=\sum_j \Tr \left[  {\mE}_j \left( \rho_p^{\lambda} \right) \rho_{q|j} ^{1-\lambda}\right]  = \sum_j \Tr \left[ \rho_p^{\lambda} {\mE}^\ast_j \left( \rho_{q|j} ^{1-\lambda} \right) \right] \ .
\eeq
%
%On the other hand, using Eq.M.~(9) and Eq.M.~(15b), we have:
On the other hand, using Eq.~\eqref{eq:chidef} and Eq.~\eqref{eqt:Feedback_tildeP}, we have:
\bes
\begin{align}
\tilde{\chi}_{\mE^{\ast}}(-\lambda) =&  \int_{-\infty}^{\infty} dv \tilde{P}_{\mE^{\ast}}(v) e^{-\lambda v} \\
&=  \int_{-\infty}^{\infty} dv \sum_{\alpha,j,\beta} \delta (v - \tilde{V}_{\beta j \alpha}) \Tr \left[ P_{\alpha} {\mE}_j^{\ast} \left(Q_{\beta|j} \right) \right]q_{\beta|j} \\
& = \sum_{\alpha, j, \beta} (p_{\alpha}/q_{\beta|j})^{\lambda} \Tr \left[ P_{\alpha}{\mE}_j^{\ast} \left(Q_{\beta|j} \right) \right]q_{\beta|j} \\
&= \sum_j \Tr \left[ \sum_{\alpha} p_{\alpha}^{\lambda} P_\alpha {\mE}_j^{\ast} \left(\sum_{\beta} q_{\beta|j}^{1-\lambda} Q_{\beta|j} \right) \right] \\
& = \sum_j  \Tr \left[\rho_p^{\lambda}  {\mE}_j^{\ast} \left(\rho_{q|j}^{1-\lambda} \right) \right] = \chi_{ \mE }(\lambda -1 ) \ ,
\end{align}
\ees
%which is the result in Eq.M.~(17).
which is the result in Eq.~\eqref{eqt:feedback_chi}.

%%%%%%%%%%%%%%%%%%%%%%%
\section{Recovering known feedback control results} \label{app:E}
%%%%%%%%%%%%%%%%%%%%%%%

We can recover known feedback control cases \cite{2010PhRvL.104i0602S,springerlink:10.1007/s10955-011-0153-7} as follows. Assume that the evolution is unitary and the intermediate measurement is error-free.  In this case, the $j$th CP map is given by
\begin{equation}
\mE_j(X) = U_j Q_j U X U^{\dagger} Q_j U_j^{\dagger} \ .
\end{equation}
This leads via Eq.~\eqref{eq:gam-feed} to 
\bes
\begin{align}
\gamma &= \sum_j \Tr \left[ \rho_{q|j} \mE_j (\ident) \right] = \sum_j \Tr \left[ \rho_{q|j} U_j Q_j  U_j^{\dagger} \right]  \\
&= \sum_j \Tr \left[Q_j U_j^{\dagger} \rho_{q|j} U_j Q_j \right] \ .
\end{align}
\ees
If we now pick the fiducial initial state $\rho$, the measurements $\mP$ and $\mQ'_j$, and the distribution $q_j$ so that they generate the Gibbs distributions
\begin{equation}
p_{\alpha} = \frac{e^{-\upbeta \varepsilon_{\alpha}}}{Z(0)} \ , \quad q_{\beta|j} = \frac{e^{-\upbeta \varepsilon_{\beta}^j}}{Z_j(t_f)} \ ,
\label{eq:pq-cont}
\end{equation}
where $\rho_{q|j} = \exp(-\upbeta H_j(t_f))/Z_j(t_f)$ is the Gibbs state associated with $H_j(t_f)$, then we recover the result for $\gamma$ derived in Ref.~\cite{springerlink:10.1007/s10955-011-0153-7}.  

We next consider the case of a classical measurement error. In this scenario when a projection on $j$ is made, there is a measurement error that gives $j'$ with probability $p_{j'|j}$, so that the feedback operation applies $U_{j'}$ instead of $U_{j}$.  The CP map is then given by:
\begin{equation}
\mE_{j'}(X) =\sum_{j} p_{j'|j} U_{j'} Q_j U X U^{\dagger} Q_j U_{j'}^{\dagger} \ .
\end{equation}
Choosing the same $p$ and $q$ distributions as in Eq.~\eqref{eq:pq-cont} with $j$ replaced by $j'$, results in
\bes
\begin{align}
\gamma &= \sum_{j'} \Tr \left[ \rho_{q|j'} \mE_{j'} (\ident) \right] = \sum_{j,j'} p_{j'|j}\Tr \left[ \rho_{q|j'} U_{j'} Q_j  U_{j'}^{\dagger} \right]  \\
&= \sum_{j,j'} p_{j'|j}  \Tr \left[Q_j U_{j'}^{\dagger} \rho_{q|j'} U_{j'} Q_j \right] \ ,
\end{align}
\ees
which is again exactly the result for $\gamma$ derived in Ref.~\cite{springerlink:10.1007/s10955-011-0153-7}.

We are free to choose a different generalized thermodynamic observable:
\begin{equation}
V_{\alpha j j' \beta} = \ln\left( p_{\alpha} /q_{\beta|j'} \right) + I_{jj'} =: -\tilde{V}_{\beta j' j \alpha}\ ,
\end{equation}
where 
\begin{equation}
I_{j j'} := \ln \left( p_{j'|j} / p_{j'} \right) \ ,
\end{equation}
the mean of which is the classical mutual information. Note that
\beq
\sum_{\beta, j'} p_{j'} q_{\beta|j'} = 1 \ .
\eeq
The joint distribution can be decomposed using Bayes' rule as $p_{(\alpha, j, j', \beta)} = p_\alpha p_{(j, j', \beta) | \alpha }$, where the conditional probability is $p_{(j, j', \beta) | \alpha} = p_{j'|j} \Tr \left[Q_{\beta|j'} U_{j'} Q_j U P_\alpha U^{\dagger} Q_j U_{j'}^{\dagger}\right]$.

We then have
\bes
\label{eqt:FeedbackError}
\begin{align}
P_{\{\mE_{j'}\}}(v)  &=& \sum_{\alpha, j,j' \beta} \delta(v - V_{\alpha j j' \beta}) p_{(\alpha, j, j', \beta)}  \\
&=& \sum_{\alpha, j,j' \beta} \delta(v - V_{\alpha j j' \beta}) p_\alpha p_{j'|j} \Tr \left[U^{\dagger} Q_j U_{j'}^{\dagger} Q_{\beta|j'} U_{j'} Q_j U P_{\alpha} \right]  \ ,
\end{align}
\ees
so that
\beq
\label{eqt:FeedbackError1}
\hspace{-1cm} P_{\{\mE_{j'}\}}(v) e^{-v} =  \sum_{\alpha, j,j' \beta} \delta(-v - \tilde{V}_{\beta j' j \alpha})q_{\beta|j'} p_{j'}  \Tr \left[U^{\dagger} Q_j U_{j'}^{\dagger} Q_{\beta|j'} U_{j'} Q_j U P_{\alpha} \right] \ .
\eeq
Note that the quantity
\begin{equation}
\sum_{\beta, j'} p_{j'} q_{\beta|j'} U_{j'}^{\dagger} Q_{\beta|j'} U_{j'} =: \hat{\rho}
\end{equation}
is a density matrix.  We can define a CPTP map:
\begin{equation}
\hat{\mE}(\hat{\rho}) = U^{\dagger}\left[  \sum_j Q_j \hat{\rho} Q_j \right] U \ .
\end{equation}
We thus recover the integrated fluctuation theorem:
\begin{eqnarray}
\int d v P_{\mE}(v) e^{-v} &=& \sum_{\alpha}  \Tr \left[P_{\alpha} \hat{\mE}(\hat{\rho}) \right] = \Tr \left[ \hat{\mE}(\hat{\rho})  \right]  =  1 \ ,
\end{eqnarray}
which is again the result in Ref.~\cite{springerlink:10.1007/s10955-011-0153-7}.
%%%%%%%%%%%%%%%%%%%%%%%
\section{Experimental system} \label{app:H}
%%%%%%%%%%%%%%%%%%%%%%%

\begin{figure}[th] %  figure placement: here, top, bottom, or page
   \centering
   \includegraphics[width=2in]{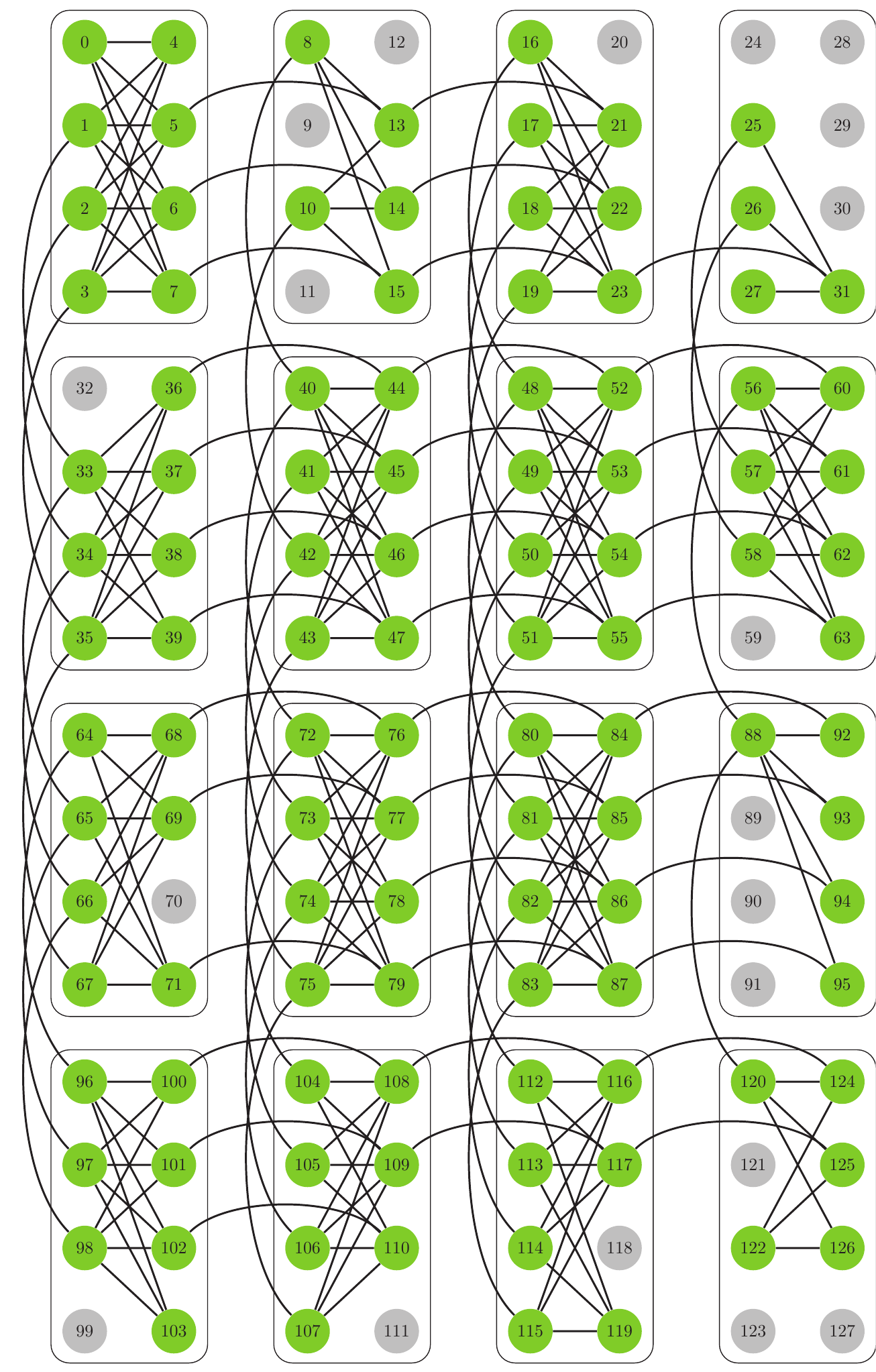} 
   \caption{A schematic of the architecture of the D-Wave One Rainier chip.  The qubits (the labelled circles) are arranged in $4 \times 4$ unit cells, with 8 qubits per unit cell.  The allowed couplings, shown by lines connecting qubits, are programmable inductive couplers.  Only green qubits corresponding to calibrated qubits are used in the experiments.}
   \label{fig:DW1}
\end{figure}

\begin{figure}[th] %  figure placement: here, top, bottom, or page
   \centering
   \includegraphics[width=3in]{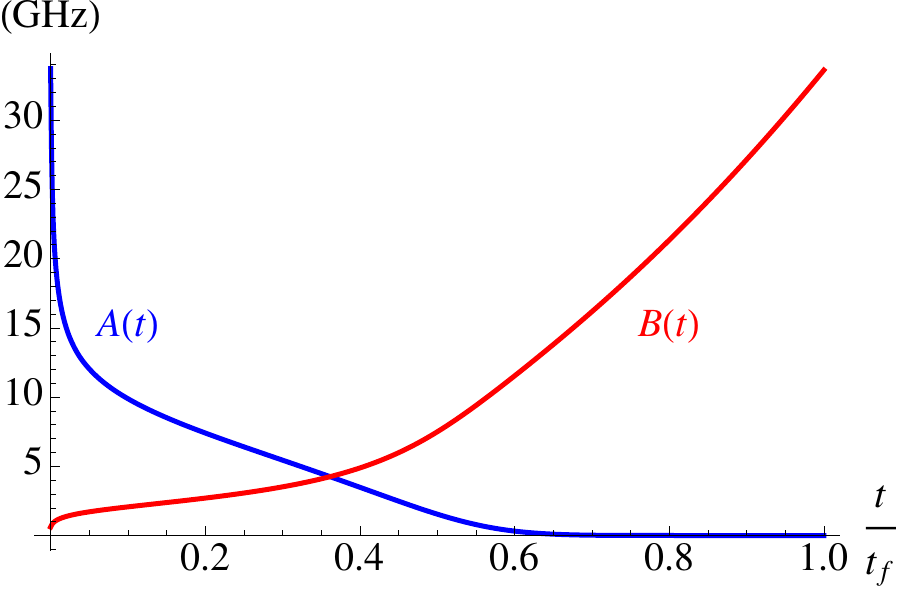} 
   \caption{The annealing schedules  $A(t)$ and $B(t)$ used in the system Hamiltonian in the experiment and in our numerical simulations.}
   \label{fig:A_B}
\end{figure}

Our experiments were performed using the D-Wave One Rainier chip at the USC Information Sciences Institute, comprising 16 unit cells of 8 superconducting flux qubits each, with a total of 108 functional qubits.   The couplings are programmable superconducting inductances.  Fig.~\ref{fig:DW1} is a schematic of the device, showing the allowed couplings between the qubits which form a ``Chimera'' graph\cite{Choi:2008,Choi:2011}.  The qubits and unit cell, readout, and control have been described in detail elsewhere \cite{Harris:2010kx,Berkley:2010zr,Johnson:2010ys}. The processor performs a quantum annealing protocol to find the ground state of a classical Ising Hamiltonian, as described by the transverse Ising Hamiltonian in Eq.~\eqref{eqt:QA_H}.  The initial energy scale for the transverse field is 33.7GHz (the $A$ function in Fig.~\ref{fig:A_B}), the final energy scale for the Ising Hamiltonian (the $B$ function) is $33.6$GHz, about $15$ times the experimental temperature of $17$mK $\approx 2.3$GHz.  The processor is programmed by specifying the qubits and the coupling strengths between qubits via a user interface.  For a given Ising Hamiltonian, the quantum annealing process was repeated $50,000$ times per qubit pair. We used five different qubit pairs to rule out systematic local magnetic field bias. As a further precaution against systematic bias we applied three spin-inversion transformations to our Ising Hamiltonian: $H_{\mathrm{Ising}} \mapsto \sigma^x_j H_{\mathrm{Ising}} \sigma^x_j$ for $j=1,2$, and $H_{\mathrm{Ising}} \mapsto \sigma^x_1 \sigma^x_2 H_{\mathrm{Ising}} \sigma^x_2 \sigma^x_1$, all of which commute with the transverse field component $\sum_i \sigma^x_i$ of our system Hamiltonian.  These transformations simply relabel the energy spectrum, i.e., if a certain spin configuration has energy $E$, then under the transformation $\sigma^x_j H_{\mathrm{Ising}} \sigma^x_j$, the configuration with the $j$-th spin flipped will also have energy $E$.  Averaging the results over the four different isospectral Ising Hamiltonians and over the five different qubit pairs for a given Ising problem, we have a total of $10^6$ data points per given values of $J$ and $t_f$.

At $J=0$, we would expect to find the (excited) states $\ket{\!\uparrow \downarrow}$ and $\ket{\!\downarrow \uparrow}$ with equal probability.  For completeness, we note that this symmetry is broken in our experimental data.  This should not be interpreted as being solely due to a local magnetic field bias, since averaging over spin-inversion transformations should have cancelled any such bias.  This suggests a more systematic (unaccounted) source of noise in the experiment.  Nevertheless, this effect does not effect our results significantly since the excited states appear very infrequently (only $\sim 7\times 10^3$ out of $5 \times 10^5$ data points for a given pair of qubits) and we find a good fit with our master equation (where this symmetry is preserved), as shown in Fig.~\ref{fig:AverageW_Experiment}.

The theoretical best fit (in Fig.~\ref{fig:AverageW_Experiment}) was found by determining the value of $\kappa = g^2 \eta / \hbar^2$ that minimizes the mean square deviation (MSD) between the $n$ experimental $\{ \langle v \rangle_{\mathrm{Ex},i} \}$ and theoretical $\{ \langle v \rangle_{\mathrm{Th},i} \}$ results:
\beq
\mathrm{MSD}(\kappa) = \frac{1}{n} \sum_{i=1}^n \left( \langle v \rangle_{\mathrm{Ex},i} - \langle v(\kappa) \rangle_{\mathrm{Th},i} \right)^2 \ .
\eeq
In principle, the high-frequency cut-off $\omega_c$ is also a free parameter in our theoretical model, and it can also be used as part of the fitting parameters.  We found that choosing a different $\omega_c$ requires a different optimal $\kappa$ value to fit the data, but we restricted ourselves to $\omega_c = 8 \pi$ GHz since it nicely satisfies the approximations made in the derivation of the master equation \cite{2012arXiv1206.4197A}.

%%%%%%%%%%%%%%%%%%%%%%%
\section{Details about the Master Equation} \label{app:F}
%%%%%%%%%%%%%%%%%%%%%%%

The underlying model assumes a total Hamiltonian of the form $H(t) = H_S(t) + H_B + H_I$, where $H_B$ is the bath Hamiltonian and $H_I$ is the system-bath interaction.
We consider a simple harmonic oscillator bath $H_B = \sum_i \sum_{k} \omega_k \left(b_k^i \right) ^{\dagger} b_k^i $, with the interaction given by the standard independent dephasing model \cite{RevModPhys.59.1}, $H_I =g \sum_{i} \sigma^z_i \otimes \sum_{k} \left( b_k^i + \left(b_k^i \right)^{\dagger} \right)$.  

The master equation used in our simulations is derived in Ref.~\cite{2012arXiv1206.4197A} and is given by (in units of $\hbar = 1$) :
\begin{eqnarray}
\hspace{-1.5cm}\dot{\rho}_S(t) &=&   - i \left[ H_S(t) + H_{\mathrm{LS}}(t), \rho_S(t) \right] \nonumber \\
&& \hspace{-1.5cm}+ \sum_{\alpha, \beta} \sum_{\omega} \gamma_{\alpha \beta} (\omega)  \left( L_{\omega,\beta}(t) \rho_S(t) L_{\omega,\alpha}^{\dagger}(t) - \frac{1}{2} \left\{L_{\omega, \alpha}^{\dagger}(t) L_{\omega,\beta}(t), \rho_S(t) \right\}\right)  \ , \label{eqt:ME}
\end{eqnarray}
where $H_{\mathrm{LS}}$ is the Lamb shift, the $\gamma$'s are dephasing and relaxation/excitation rates.  The Lindblad operators are given by:
\begin{equation}
L_{\omega,\alpha}(t) = \sum_{\omega = \varepsilon_b(t) - \varepsilon_a(t)} | \varepsilon_a(t) \rangle \langle \varepsilon_a(t) |  \sigma^z_{\alpha} | \varepsilon_b(t) \rangle \langle \varepsilon_b(t) | \ ,
\end{equation}
where the instantaneous Bohr frequency $\omega$ is expressed in terms of the instantaneous energy eigenstates, i.e., $H_S(t) | \varepsilon_a(t) \rangle = \varepsilon_a(t)  | \varepsilon_a(t) \rangle$
For an Ohmic bath with high-frequency cut-off $\omega_c$, we have
\begin{eqnarray}
\gamma_{\alpha \beta}(\omega) &=& \delta_{\alpha, \beta} \frac{g^2 \eta \omega e^{-\omega / \omega_c}}{1 - e^{-\upbeta \omega} } \ , \quad
 H_{\mathrm{LS}} = \sum_{\alpha \beta} \sum_{\omega} S_{\alpha \beta}(\omega) L_{\omega,\alpha}^{\dagger}(t) L_{\omega,\beta}(t)  \ ,  \nonumber \\
S_{\alpha \beta}(\omega) &=& \int_{-\infty}^{\infty} \frac{ d \omega'}{2 \pi} \gamma_{\alpha \beta}(\omega') \mathcal{P} \left(\frac{1}{\omega - \omega' } \right) \ .
\end{eqnarray}
where $\eta$ is a parameter (with dimension time squared) characterizing the Ohmic bath, and $\mathcal{P}$ denotes the Cauchy principal value.  To see that this CP map is not unital, we plug $\ident$ into the RHS of Eq.~\eqref{eqt:ME} and find that the non-zero component arises from the dissipative part associated with relaxation and excitation processes
\begin{equation} \label{eqt:non-unital}
\sum_{\alpha, \beta} \sum_{\omega} \gamma_{\alpha \beta}(\omega) \left[ L_{\omega, \beta}(t), L_{\omega,\alpha}^{\dagger}(t) \right] \ .
\end{equation}
Since $\gamma_{\alpha \beta}(\omega) \neq \gamma_{\beta \alpha}(-\omega)$, meaning that the relaxation and excitation transition rates are unequal, the term \eqref{eqt:non-unital} is non-zero, making the CPTP map generated by the master equation non-unital.  
%\red{Do we want to make the SCL comment here, that for SCL the transition and relaxation rates are equal hence why it is unital?}

%%%%%%%%%%%%%

\section{Numerical confirmation of the QJE and first moment expression for the adiabatic Markovian master equation} \label{app:G}
\begin{figure}[t] %  figure placement: here, top, bottom, or page
   \centering
   \includegraphics[width=3in]{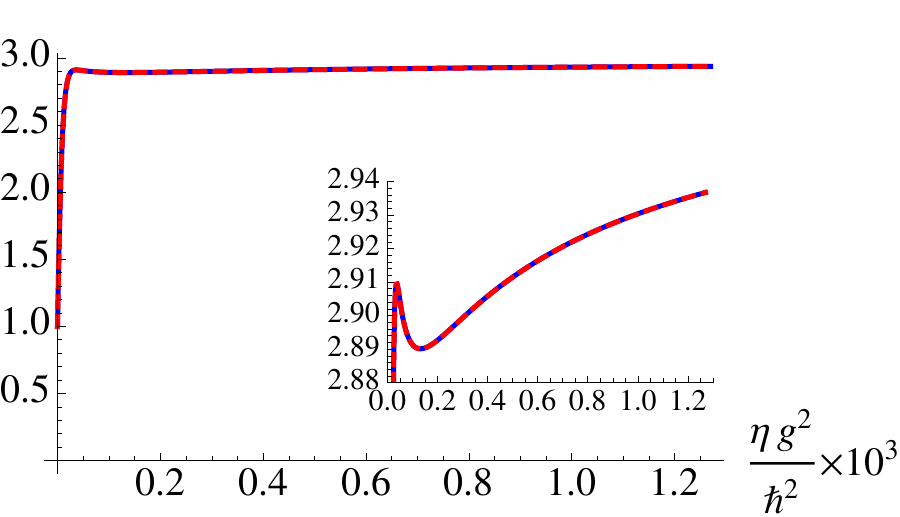} 
   \caption{Independent calculation of  $\langle e^{-\upbeta \left( \Delta E - \Delta F \right)} \rangle$ (blue solid) and $\gamma$ (red dashed) as a function of the system-bath coupling strength using the adiabatic Markovian master equation with $t_f = 5 \mu s$, $J=1/2$, $T = 17$mK, and $\omega_c = 8 \pi$.  Inset: magnification of the behavior near the top of the curves. The two curves overlap to within numerical accuracy.}
   \label{fig:Lindblad_Exp_vs_Gamma}
\end{figure}

\begin{figure}[t] %  figure placement: here, top, bottom, or page
   \centering
   \includegraphics[width=3in]{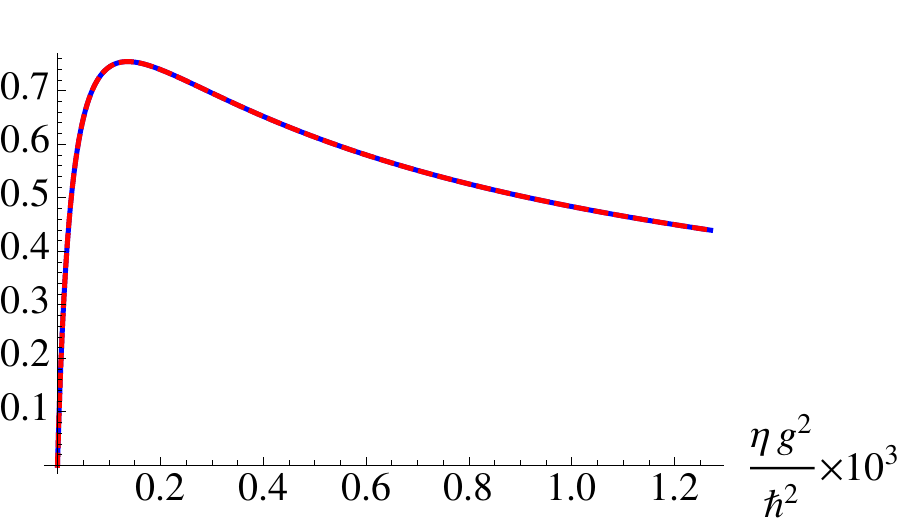} 
   \caption{Simulation results for $\langle \upbeta \left( \Delta E - \Delta F \right) \rangle$ (solid blue) and the RHS of Eq.~(13a) (red dashed) using the 
   Markovian master equation and $t_f = 5 \mu s$, $J=1/2$, $T = 17$mK, and $\omega_c = 8 \pi$. The two curves overlap to within numerical accuracy.}
   \label{fig:AverageW_Lindblad_tf=5}
\end{figure}

%In order to test Eq.M.~(18) and Eq.M.~(13a) we performed the following simulations for the two-qubit model 
In order to test Eq.~\eqref{eqt:Specific} and Eq.~\eqref{eqt:Paolo} we performed the following simulations for the two-qubit model described in the main text.  We initialized the system in one of the four energy eigenstates, $\rho_S(0) = |\varepsilon_a(0)\rangle \langle \varepsilon_a(0)|$, then we evolved the density matrix using our adiabatic master equation.  The diagonal elements of the density matrix at $t = t_f$ are then associated with the probability $p_{\beta | \alpha}$ of measuring the state $|\varepsilon_\beta(t_f) \rangle$.  Using this we calculated the expectation value $\av{ e^{-\upbeta \left( \Delta E - \Delta F \right)} }$.  We then used our adiabatic master equation to evolve the identity operator.  This allowed us to numerically find $\mE(\ident)$, which in turn allowed us to calculate $\gamma$.  %The equality expressed in Eq.M.~(18) is obtained with high accuracy, as shown in Fig.~\ref{fig:Lindblad_Exp_vs_Gamma}.
The equality expressed in Eq.~\eqref{eqt:Specific} is obtained with high accuracy, as shown in Fig.~\ref{fig:Lindblad_Exp_vs_Gamma}.
%
%We can also calculate the LHS and RHS of the first moment of our fluctuation theorem [Eq.M.~(13a)] i
We can also calculate the LHS and RHS of the first moment of our fluctuation theorem [Eq.~\eqref{eq:S-proj}] independently.  We again find excellent agreement between the two results: see Fig.~\ref{fig:AverageW_Lindblad_tf=5}.

%We can try something similar but now with a master equation in the singular coupling limit (see ref.~\cite{} for details).  In this case, the map is unital, and therefore, we expect $\gamma=1$.  We can confirm this, as shown in Fig.~\ref{fig:SCL_Exp_vs_\mGamma}.
%
To compute $\Delta F$, as needed for Eq.~\eqref{eq:<v>}, the eigenvalues $\varepsilon_i$ of the initial ($i=\alpha$) and
final ($i=\beta$) Hamiltonian are numerically computed by diagonalizing the respective Hamiltonians.  In turn, the respective partition functions are calculated using the energy eigenvalues found, $Z = \sum_i e^{-\beta \varepsilon_i}$.  The free energy is then $F = -\ln(Z)/\beta$.

%%%%%%%%%%%%%%%%%%%%%%%%
%\section{Hamiltonian Gauge Freedom}
%%%%%%%%%%%%%%%%%%%%%%%%%
%%
%
%There is a gauge freedom in choosing an overall constant for any Hamiltonian.  The choice of gauge does not change the result of the dynamics, so our results so far should be invariant under this gauge.  This can be demonstrated as follows.  Let us consider a (possibly time dependent) shift of the Hamiltonian:
%%
%\beq
%H'(t) = H(t) + a(t)\ident\, .
%\eeq
%%
%The shift scales the partition function $Z'(t) = e^{-\upbeta a(t)} Z(t)$ and hence shifts the free energy $F'(t) = F(t) + a(t)$ (and all energies $\eps'(t) = \eps(t) + a(t)$).  However, the Gibbs state remains invariant under this shift.  Using these observations, the observable $V_{\alpha \beta}$ is invariant under the shift:
%%
%\bes
%V_{\alpha \beta}' &= & \upbeta \left( \eps'_{\beta}(t_f) - \eps'_{\alpha}(0) - F'(t_f) + F'(0) \right) \\
%&= & \upbeta \left (\eps_{\beta}(t_f) - \eps_{\alpha}(0) - F(t_f) + F(0) \right) = V_{\alpha \beta}
%\ees
%%
%Therefore, our results remain invariant under different gauge choices.

%%%%%%%%%%%%%%%%%%%%%%%
%\section*{References}
%%%%%%%%%%%%%%%%%%%%%%%%
%\bibliographystyle{unsrt}
%\bibliography{refs2}

\end{document}